\documentclass{aa}
\usepackage[utf8]{inputenc}
\usepackage[varg]{txfonts}
\usepackage[english]{babel}

\usepackage{graphicx}

\usepackage[tight]{subfigure}

\usepackage[table,svgnames]{xcolor}

\usepackage{natbib}

\newcommand{\eturb}{\ensuremath{\epsilon_{\rm turb}}}

\title{The influence of turbulence during magnetized core collapse and its consequences on low-mass star formation}
\titlerunning{Influence of turbulence on low-mass star formation}
\author{Marc \textsc{Joos}\inst{\ref{ens},\ref{cea}} \and Patrick \textsc{Hennebelle}\inst{\ref{ens},\ref{cea}} \and Andrea \textsc{Ciardi}\inst{\ref{ens}} \and Sebastien \textsc{Fromang}\inst{\ref{cea}}}
\authorrunning{Marc \textsc{Joos} et al.}
\institute{Laboratoire de radioastronomie, LERMA, Observatoire de Paris, \'Ecole Normale Supérieure, Université Pierre et Marie Curie (UMR 8112 CNRS), 24 rue Lhomond, 75231 \textsc{Paris} Cedex 05, \textsc{France}\label{ens} \and CEA, IRFU, SAp, Centre de Saclay, 91191 \textsc{Gif-Sur-Yvette}, \textsc{France} \label{cea}}

\abstract{Theoretical and numerical studies of star formation have shown that a magnetic field can greatly influence both disk formation and its fragmentation, with even relatively low magnetic field strengths being able to prevent these processes. However, very few studies have investigated the combined effects of a magnetic field and turbulence.} 
{We study the collapse of turbulent, magnetized prestellar cores, focusing on the effects of magnetic diffusion, and misalignment between rotation axis and magnetic field, on the formation of disks, fragmentation and the generation of outflows.} 
{We perform three-dimensional, adaptive mesh, numerical simulations of magnetically super-critical collapsing dense cores of 5~$M_{\odot}$ using the magneto-hydrodynamic code \textsc{Ramses}. A \emph{turbulent} velocity field is imposed as initial conditions, characterised by a Kolmogorov power spectrum. Different levels of turbulence (a laminar case as well as subsonic and supersonic cases) and magnetization (from weak to strong magnetization) are investigated, as well as three realisations for the turbulent velocity field.} 
{The turbulent velocity field imposed as initial conditions contains a non-zero angular momentum, which is responsible for a misalignment of the rotation axis with respect to the initial magnetic field, and an effective turbulent diffusivity in the vicinity of the core. Both effects are responsible for a significant decrease of the magnetic braking, and facilitate the formation of early massive disks. These disks can fragment even with $\mu\sim5$ at late times, in contrast to simulations of 1~$M_{\odot}$ cores, where fragmentation is prevented for such values of $\mu$. Slow asymmetric outflows are always launched, and they carry a mass which is comparable to that of the adiabatic first core.}
{Because of turbulence-induced misalignment and magnetic diffusivity, massive disk formation is possible; nevertheless their mass and size are much reduced compared to those of disks formed in unmagnetized collapsing cores. We find that for $\mu\geq5$ fragmentation can occur.} 


\begin{document}

\maketitle

\section{Introduction}

Disks are a key element in star formation: protostars probably grow by accreting material from protostellar accretion disks \citep{Larson03}, which at later stages are the natural progenitors of planets \citep{Lissauer93}. The formation of disks is a challenging theoretical question, and the conditions for their presence remain unclear. Magnetic fields certainly play an important role, and can efficiently transport angular momentum away from the central regions of the collapsing core, preventing the early formation of massive disks. This occurs even for relatively low magnetic field intensities \citep{Mellon08, Price07, Hennebelle08a}, at a level largely in accordance with observational measurement of core magnetizations\citep{Crutcher99,Falgarone08}. However, this so-called magnetic braking ``catastrophe'' may in fact be compatible with observations. Unlike Class I and more evolved sources, the presence of disks around Class 0 sources is not well established, and recent studies show no clear evidence of disks or fragmentation in Class 0 protostars \citep{Maury10}.

 Several studies have investigated non-ideal magnetohydrodynamics (MHD) effects on the magnetic braking and on disk formation. While \cite{Mellon09,Li11} found that ambipolar diffusion does not allow the building of massive disks, the effects Ohmic dissipation remain less clear, with only small disks able to form\citep[$\sim$ 10~AU][]{Dapp10}. Other means of redistributing magnetic flux were also investigated. \cite{Krasnopolsky12} found that flux expulsion from the inner core regions by both ambipolar and Ohmic diffusion, can promote the development of an interchange which is very effective in advecting outward the magnetic field, nevertheless they did not observe disks.

The effects of turbulence were investigated by \cite{Santos-Lima12} who found that an effective turbulent diffusivity \citep[of the same order of magnitude as the enhanced resistivity of][]{Krasnopolsky10} is sufficient to decreases the magnetic braking efficiency, and generate disks. Similarly, in the context of massive star formation, \cite{Seifried12} argued the lack of coherent rotational velocity field in their turbulent simulations prevents the build up of a toroidal magnetic field component, and reduces the magnetic braking. However, they also noted that turbulence induces a strong misalignment between the rotation axis of the disk and the magnetic field. This misalignment was shown in fact to be critical in the transport of angular momentum by \cite{Hennebelle09} and \cite{Joos12}. In ideal MHD, \cite{Hennebelle09} and \cite{Joos12} investigated the influence of the orientation of the rotation axis with respect to the magnetic field on disk formation, showing both analytically and numerically that the tilt of the rotation axis can reduce significantly the magnetic braking, allowing the formation of massive disks.

Understanding the formation of multiple systems is another crucial issue related to the angular momentum distribution. A large fraction of stars are actually observed to be in binary or higher multiple systems \citep{Duquennoy91}. The question of the multiple system formation was addressed by several authors using numerical simulations. While the magnetic field is a key ingredient of star formation theory, simulations investigating multiple star formation were until recently mainly hydrodynamical. Fragmentation of collapsing cores is thought to be the most plausible process to form multiple systems. Pioneer work from \cite{Boss79} already suggested that binaries could form from fragmentation of the collapsing cloud. The robustness of these results, as well as the initial parameters of the collapse (as the initial shape of the cloud), were thoroughly investigated \citep[see for example][]{Bonnell91,Bate95}. Some authors took into account the effect of turbulence \citep{Goodwin04a,Goodwin04b}, showing that binary and multiple system can form even with low levels of turbulence, and without a large initial amount of rotational energy. Fragmentation of the protostellar disk was also studied \citep{Stamatellos07}, illustrating that companion stars could form in an extended massive disk around the protostar. The effects of radiative transfer, without \citep{Offner09} or with \citep{Offner10} turbulence were also addressed. The main conclusion being that while radiative transfer inhibits fragmentation in the disk, turbulence tends to favour fragmentation of the cloud.

Few studies have investigated the influence of a magnetic field on the fragmentation process, and they reveal in particular that it can very efficiently prevent fragmentation \citep{Hennebelle08b,Machida05}. The magnetic pressure has a stabilizing influence, particularly because of the growth of the toroidal component of the magnetic field which is responsible for disk stabilization. Radiative processes do not enhance fragmentation in the presence of a magnetic field \citep{Commercon10}, on the contrary, the more realistic heating of the disk inhibits the fragmentation even further. Fewer studies investigated the role of turbulence in magnetized low-mass cores, as for example in the work of \cite{Matsumoto11}, whose simulations, although not really focused on fragmentation itself, showed no signs of it.

The goal of this paper is to study in detail disk formation and fragmentation in the presence of both a magnetic field and turbulence. The effects of turbulence highlighted in the present work are twofold. First, turbulence generates an effective magnetic diffusivity by promoting magnetic reconnection \citep{Weiss66,Santos-Lima12}, which helps to expel the magnetic flux from the central regions of the core. Second, it is responsible for a ``local'' misalignment between the rotation axis and the magnetic field lines, which results in a significant decrease of the magnetic braking efficiency, as discussed in \cite{Joos12}.

The plan of the paper is the following. In section 2, we present our numerical setup and initial conditions. The effects of turbulence are presented in section 3: the diffusion of the magnetic field, the impact on the orientation of the angular momentum with respect to the magnetic field, and its implications on the magnetic braking efficiency. The consequences on star formation are presented in section 4: the effects on disk formation, fragmentation, and outflows. In section 5, we discuss the influence of misalignment on a laminar case and present results for different realisations of the turbulent velocity field. Section 6 concludes the paper.

\section{Numerical setup and initial conditions}
 
\subsection{Numerical setup}

We perform 3D numerical simulations with the AMR code \textsc{Ramses}~\citep{Teyssier02, Fromang06a}. \textsc{Ramses} can treat ideal MHD problems with self-gravity and cooling. The magnetic field evolves using the constrained transport method, preserving the nullity of the divergence of the magnetic field. The high resolution needed to investigate the problem is provided by the AMR scheme. Our simulations are performed using the HLLD solver \citep{Miyoshi05}.

The calculations start with 128$^{3}$ grid cells. As the collapse proceeds, new cells are introduced to ensure the resolution of the Jeans length with at least ten cells. Altogether we typically use 10 AMR levels during the calculation, providing a finest spatial resolution of $\sim$ 0.4~AU. 

\subsection{Initial conditions}

We consider simple initial conditions consisting of a spherical cloud with a mass of 5~$M_{\odot}$. Compared to the one solar mass clouds usually studied in the context of low-mass star formation, more massive clouds are expected to be more turbulent \citep{Larson81}. The density profile of the initial prestellar core is given by
\begin{equation}
\rho=\frac{\rho_{0}}{1+(r/r_{0})^{2}},
\end{equation}
 with $\rho_{0}=1.2\times10^{-18}$~g~cm$^{-3}$ the central density, $r_0=2.3 \times 10^{-2}$~pc the radius of the central region of the prestellar core, where the density is approximately uniform, and $r_B=0.14$~pc the boundary of the prestellar core; these are in accordance with observations \citep{Andre00,Belloche02}. The ratio of the thermal over gravitational energy is about 0.25 whereas the ratio of the rotational over gravitational energy $\beta$ is about 0.03. The magnetic field intensity is quantified through the magnetization parameter $\mu$, defined as the mass-to-flux over the critical mass-to-flux ratio (its critical value being $\simeq 1/(2\pi G^{1/2})$). Various magnetizations are studied, corresponding to a magnetization parameter $\mu=$ 2, 5 (magnetized super-critical cloud, in agreement with observations, as pointed out in the introduction) and 17 (very super-critical cloud). The magnetic field is initially aligned with the rotation axis. We emphasize that the $\mu=5$ is an intermediate case: in the $\mu=2$ case, the magnetic field dominates the dynamics whereas the $\mu=17$ case is a quasi-hydrodynamical case. Therefore in the rest of the paper particular attention will be payed to the $\mu=5$ case.

A random phase velocity field with a Kolmogorov power spectrum is imposed as initial condition. The turbulent velocity field is a mixture of compressible and solenoidal modes generated, nearly in equipartition, using a fractional Brownian motion, following \cite{Levrier04}. The largest wavelength excited is comparable to the size of the central region of the prestellar core, $r_{B}$. We explore mostly one realisation of the velocity field, but we run two other realisations for the $\mu=5$ simulations in order to assess the robustness of our conclusions. Prestellar cores are thought to lay in a transonic turbulent environment \citep{Ward-Thompson07}. Therefore, various initial turbulent energy are studied: \eturb{} = 0, 0.1, 0.2 and 0.5, where \eturb{} is the ratio of the turbulent over gravitational energy $E_{\rm turb}/E_{\rm grav}$. These correspond to a laminar case, a subsonic case, and two supersonic cases respectively. Specifically, the values \eturb{} = 0.1 corresponds to a Mach number $\mathcal{M}=0.85$, \eturb{} = 0.2 to $\mathcal{M}=1.2$, and \eturb{} = 0.5 to $\mathcal{M}=1.9$. Table~\ref{tab:simu} lists all the simulations parameters.

\begin{figure}
\subfigure[\label{img:am0}]{\includegraphics[width=0.4\textwidth]{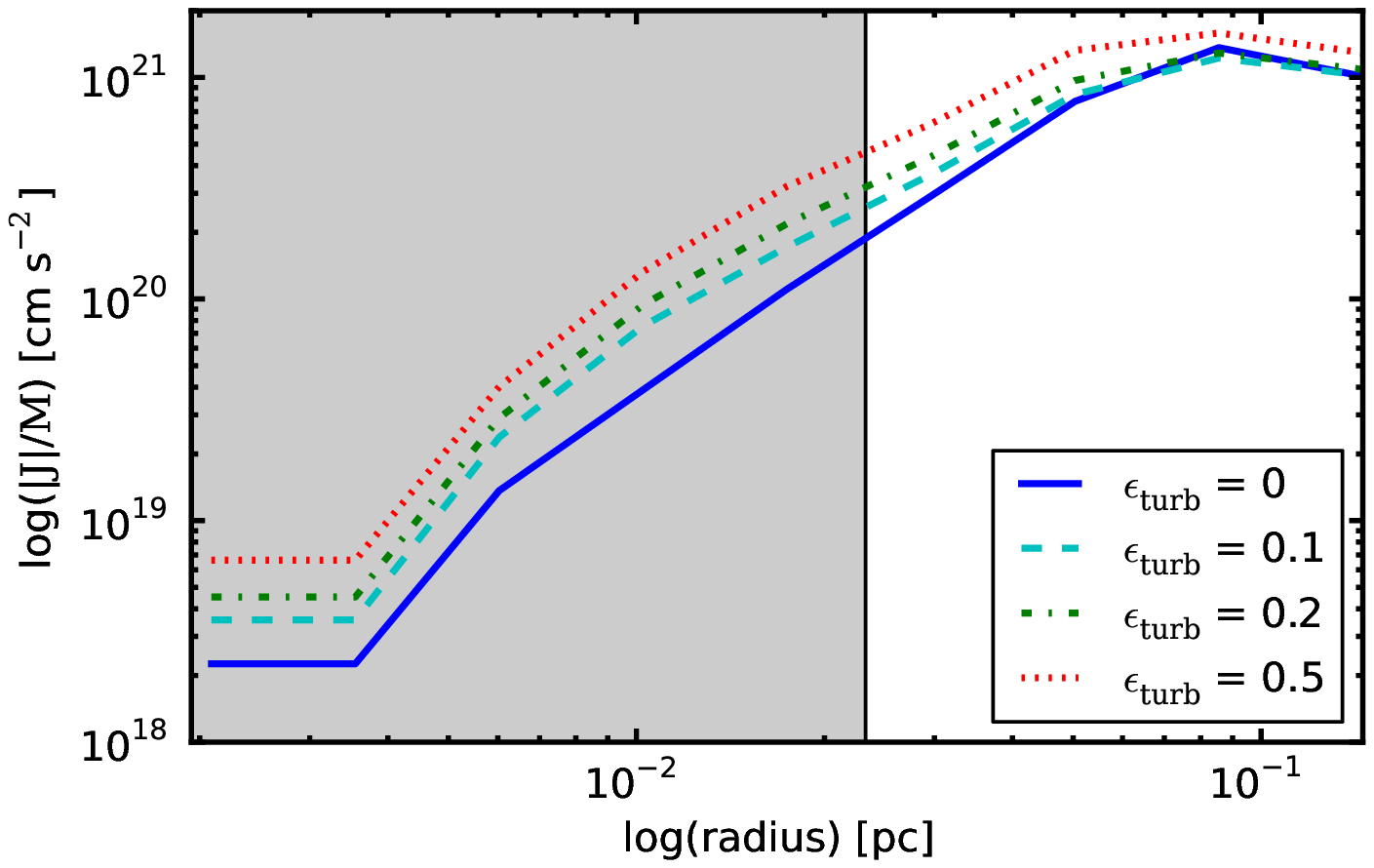}}
\subfigure[\label{img:ang0}]{\includegraphics[width=0.4\textwidth]{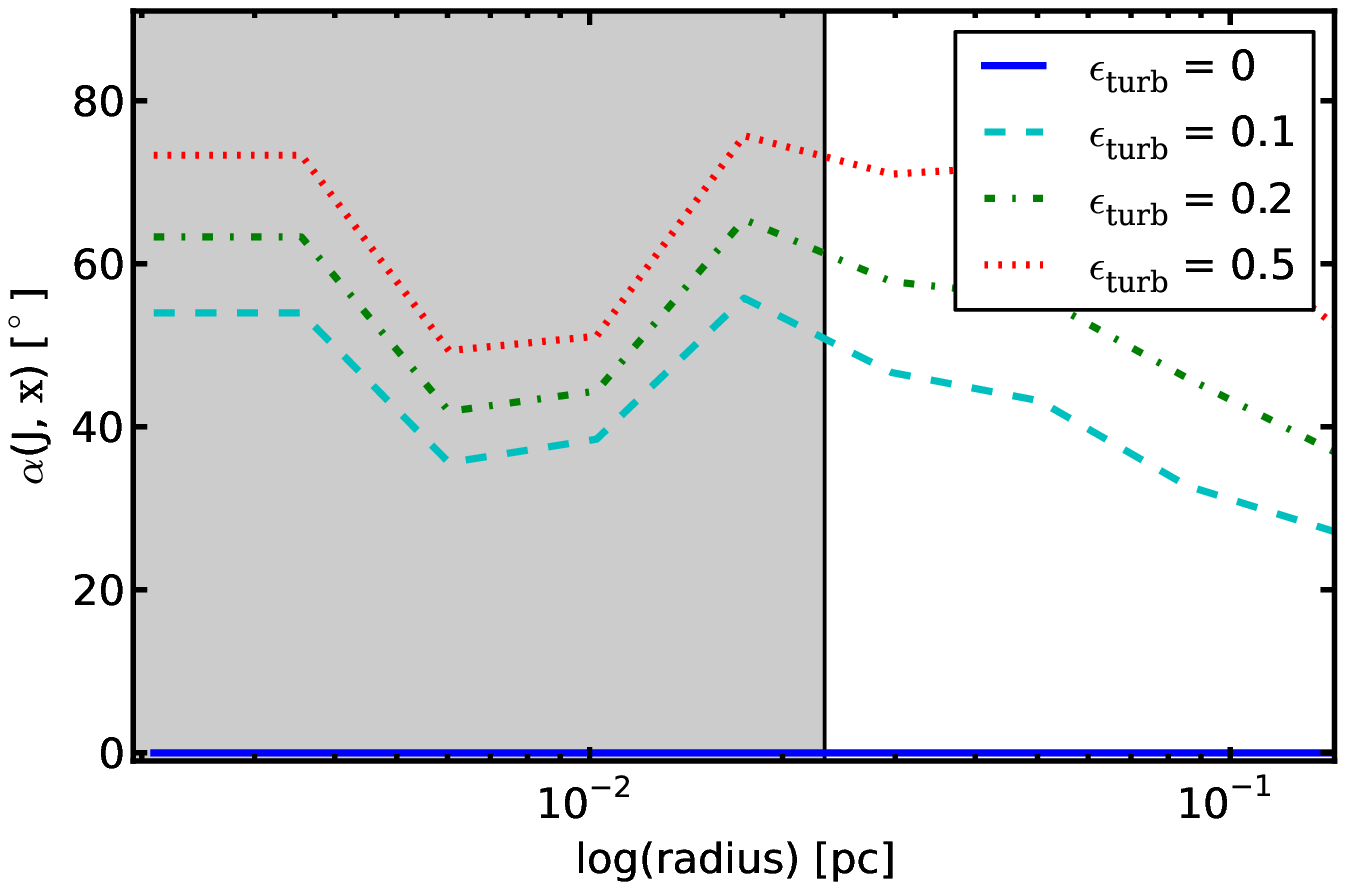}}
\caption{Specific angular momentum (upper panel) and initial angle between the rotation axis and the $x$-axis at $t=0$~yr. The shaded area corresponds to the central region of the prestellar core of radius $r_0$.} 
\label{img:turbini} 
\end{figure}

In addition to the angular momentum imposed initially on the cloud, the turbulent velocity field can generate a supplementary non-zero angular momentum. Figure \ref{img:am0} shows that in the inner parts of the prestellar core, the specific angular momentum is larger when turbulence is present. In a sphere of radius $r_{0}$, the angular momentum is larger by a factor $\sim$1.3 to $\sim$2, and the larger the turbulent energy, the larger the angular momentum. However outside the central region of the prestellar core, the angular momentum tends to converge to a unique value, independently of the initial turbulent energy. This also implies the presence of an initial tilt of the rotation axis with respect to the $x$-axis, which is given by the initial direction of the magnetic field, as shown by Fig.~\ref{img:ang0}. With turbulence, this angle is initially larger than 30$^{\circ}$. However, due to the turbulent motions, this angle evolves rapidly after the beginning of the simulation, as discussed in the next section.

To avoid the formation of a singularity and to mimic the fact that at high density the gas becomes opaque i.e. nearly adiabatic, we use a barotropic equation of state 
\begin{equation}
\frac{P}{\rho}=c_s^2=c_{s,0}^2\left[1+\left(\frac{\rho}{\rho_{\rm ad}}\right)^{2/3}\right],
\end{equation}
 where $\rho_{\rm ad}$ is the critical density over which the gas becomes adiabatic; we take $\rho_{\rm ad}=10^{-13}$ g~cm$^{-3}$. When $\rho>\rho_{\rm ad}$ the adiabatic index $\gamma$ is therefore equal to 5/3, which corresponds to an adiabatic mono-atomic gas. At lower density, when the gas is isothermal, $P/\rho$ is constant, with $c_{s,0} \sim 0.2$ km~s$^{-1}$. The corresponding free-fall time is $t_{\rm ff} \sim 60$~kyr (for an initial central density of about $1.2 \times 10^{-18}$~g~cm$^{-3}$).

Note that for a 5~$M_{\odot}$ prestellar core, it is likely that the barotropic equation of state is not accurate enough, and that radiative transfer calculations should be employed \citep{Commercon10,Commercon11a,Commercon11b,Tomida10}. However, such calculations are computational intensive and do not allow an easy exploration of the space parameters.

\begin{table}
\centering %
\begin{tabular}{c | c c c}
\hline 
run  & $\mu$  & $\eturb=E_{{\rm turb}}/E_{{\rm grav}}$  & realisation \\
\hline 
\hline 
CT200  & 2  & 0  & -- \\
CT2100  & 2  & 0.1  & 0 \\
CT2200  & 2  & 0.2  & 0 \\
CT2500  & 2  & 0.5  & 0 \\
CT2201  & 2  & 0.2  & 1 \\
CT2501  & 2  & 0.5  & 1 \\
CT2202  & 2  & 0.2  & 2 \\
CT2502  & 2  & 0.5  & 2 \\
\hline 
CT500  & 5  & 0  & -- \\
CT5100  & 5  & 0.1  & 0 \\
CT5200  & 5  & 0.2  & 0 \\
CT5500  & 5  & 0.5  & 0 \\
CT5201  & 5  & 0.2  & 1 \\
CT5501  & 5  & 0.5  & 1 \\
CT5202  & 5  & 0.2  & 2 \\
CT5502  & 5  & 0.5  & 2 \\
\hline 
CT1700  & 17  & 0  & -- \\
CT1710  & 17  & 0.1  & 0 \\
CT17200  & 17  & 0.2  & 0 \\
CT17500  & 17  & 0.5  & 0 \\
\hline 
\end{tabular}\caption{List of performed simulations, with $\mu$ the magnetization parameter, $\eturb=E_{\rm turb}/E_{\rm grav}$ the ratio of turbulent over gravitational energies, and the realisations performed.}

\label{tab:simu} 
\end{table}

\section{Effects of turbulence}

In this section, we will discuss the main two main effects of turbulence on the collapse of a prestellar core: namely that the magnetic field can be efficiently diffused out of the inner regions of the collapsing core owing to an effective turbulent diffusivity, and that the angular momentum induced by the initial turbulent velocity field is responsible for a ``local'' misalignment between the rotation axis and the magnetic field. This two effects have important consequence on the magnetic braking, and therefore on the angular momentum distribution. 

\subsection{Magnetic diffusion}

Let us begin by investigating the diffusion of the magnetic field in the regions around the first Larson's core (the adiabatic core, which will be referred to ``first core'' in the rest of the paper).

To estimate the diffusion of the magnetic field in our simulation, we first estimate the magnetization around the first cores. All physical quantities are computed for a sphere of radius $R$; to express the magnetic flux, given by $\pi R^2\langle{\bf B}\rangle$, we consider the mean value of the magnetic field on the sphere $\langle{\bf B}\rangle$ (which is volume averaged). The mass of the sphere is given by $M(R)$ and the critical mass-to-flux ratio is defined as $(M/\Phi)_{\rm cr}=1/(2\pi\sqrt{G})$. The time evolution of the magnetization parameter $\mu(t)|_R$ of a sphere of a given radius $R$ can therefore be expressed as 
\begin{equation}
\mu(t)|_{R}\simeq\frac{M(R)}{\pi R^{2}\langle{\bf B}\rangle}\left/\frac{1}{2\pi\sqrt{G}}\right..\label{eq:mu}
\end{equation}
 We compute $\mu(t)|_R$ for $R=100,\,500$ and 1000~AU.

\begin{figure*}
  \subfigure[\label{img:diffmu17}]{\includegraphics[width=1.\textwidth]{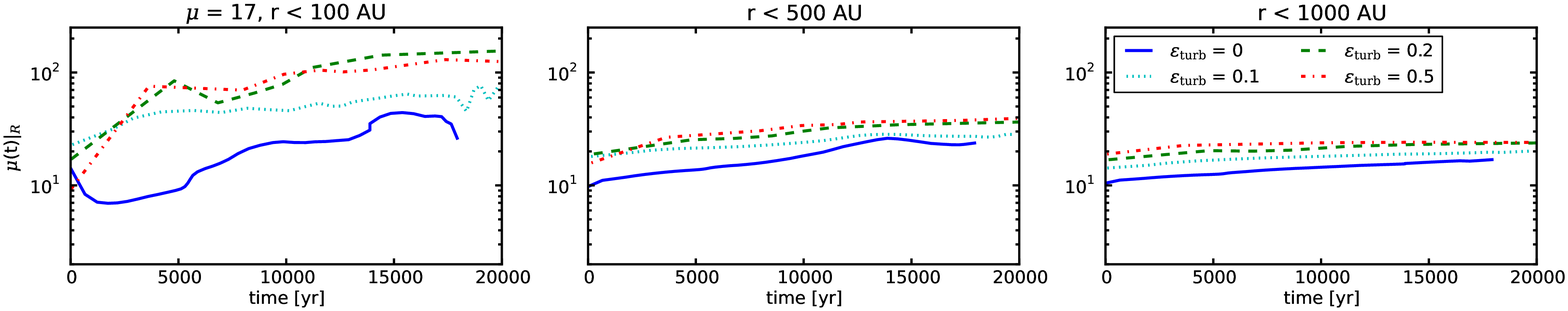}}
  \subfigure[\label{img:diffmu05}]{\includegraphics[width=1.\textwidth]{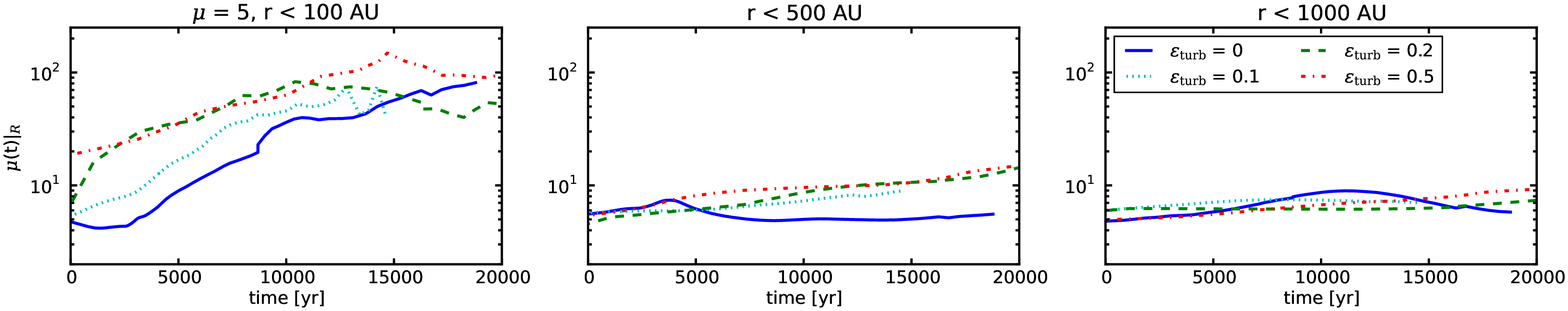}}
  \subfigure[\label{img:diffmu02}]{\includegraphics[width=1.\textwidth]{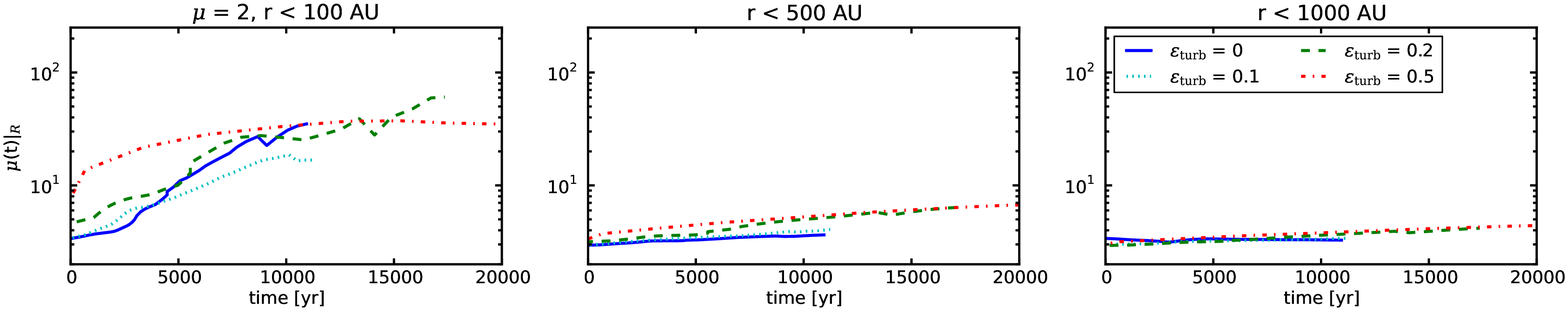}}
 \caption{Evolution of the magnetization parameter within a radius $R=$ 100, 500 and 1000~AU around the first core (left, central and right panels respectively) for an initial $\mu$ of 17, 5 and 2 (upper, middle and lower panels) and for \eturb{} = 0, 0.1, 0.2, and 0.5. $t=0$~yr corresponds to the formation of the first core.}
  \label{img:diffmu} 
\end{figure*}
 
Figure \ref{img:diffmu} shows the evolution $\mu(t)|_{R}$ for initial values of $\mu=$ 17, 5 and 2. Time $t=0$~yr corresponds to the formation of the first core. We note that this is different from \cite{Joos12}; here the choice is motivated by the fact that turbulence modifies the gas dynamics and delays the collapse. The actual delay depending on the realisation of the turbulent velocity field and on the turbulent energy.

For all magnetizations, the magnetic flux is initially well-conserved at large scale: within a sphere of 500 or 1000~AU, the estimated magnetization parameter remains approximately constant. However, at later times, the magnetic flux starts to decrease, for example for $R=500$ AU the magnetization parameter increases by a factor $\sim$2.

In the laminar case (\eturb{} = 0), the magnetic flux is first well-conserved at smaller radii (within a sphere of radius $R=100$~AU). However at times $\gtrsim$5000~yr after the formation of the first core, it starts to be diffused up to $\sim$2 to 10 times its initial value, this is likely a consequence of the interchange instability, as discussed by \cite{Krasnopolsky12}.

We find that with an initial turbulent velocity field (\eturb{} = 0.1, 0.2, and 0.5) the magnetic flux is diffused out more efficiently in the vicinity of the first core ($R\lesssim100$ AU). The larger the initial turbulent energy, the larger is$\mu$. For example, $\mu$ is typically $\sim$3 to 10 times larger than its initial value in the subsonic case (\eturb{} = 0.1), whereas it is $\sim$10 times larger than its initial value for \eturb{} = 0.2, and $\sim$10 to 30 times larger for \eturb{} = 0.5.

Following the method of \cite{Fromang06}, we have also computed an approximate estimate of the turbulent resistivity. We obtain a magnetic diffusivity of about $2 \times 10^{21}$~cm$^2$~s$^{-1}$ for \eturb{} = 0.2, and $4 \times 10^{21}$~cm$^2$~s$^{-1}$ for \eturb = 0.5. Therefore, the larger the turbulent energy, the larger the diffusivity coefficient, which is in good agreement with the measured diffusion of the magnetic field. There is roughly a factor 2 between the diffusivity coefficient for the \eturb{} = 0.2 and 0.5 cases, and consequently the magnetization parameter in the vicinity of the first core was increased by a factor two between those two turbulent cases.

We note that even a low level of turbulence is sufficient to efficiently diffuse the magnetic flux from the first core. The significant diffusion of magnetic field even at larger radii ($R=500$~AU) is crucial for the transport of angular momentum, as strong magnetic braking occurs in those regions, and we can expect it to influence the formation of protostellar disks.

\subsection{Orientation} 
\label{sec:orientation}

As we discussed in \cite{Joos12}, the orientation of the rotation axis with respect to the direction of the magnetic field has important consequences on the magnetic braking and the transport of angular momentum. One of the main conclusions was that even with a relatively strong magnetic field ($\mu \gtrsim 3$), a relatively small misalignment ($\gtrsim 20^{\circ}$), is sufficient to enable the formation of massive disks. Although those simulations were performed without turbulence, we suggested that turbulent motions in the ISM could be at the origin of this misalignment (see the section 3.1.5 of \cite{Joos12}). In the present case, the turbulent velocity field imposed as initial conditions induces an initial tilt of the angular momentum, with values of the angle laying between 20 and 60$^{\circ}$.

\begin{figure*}
  \subfigure[\label{img:angle17}]{\includegraphics[width=0.89\textwidth]{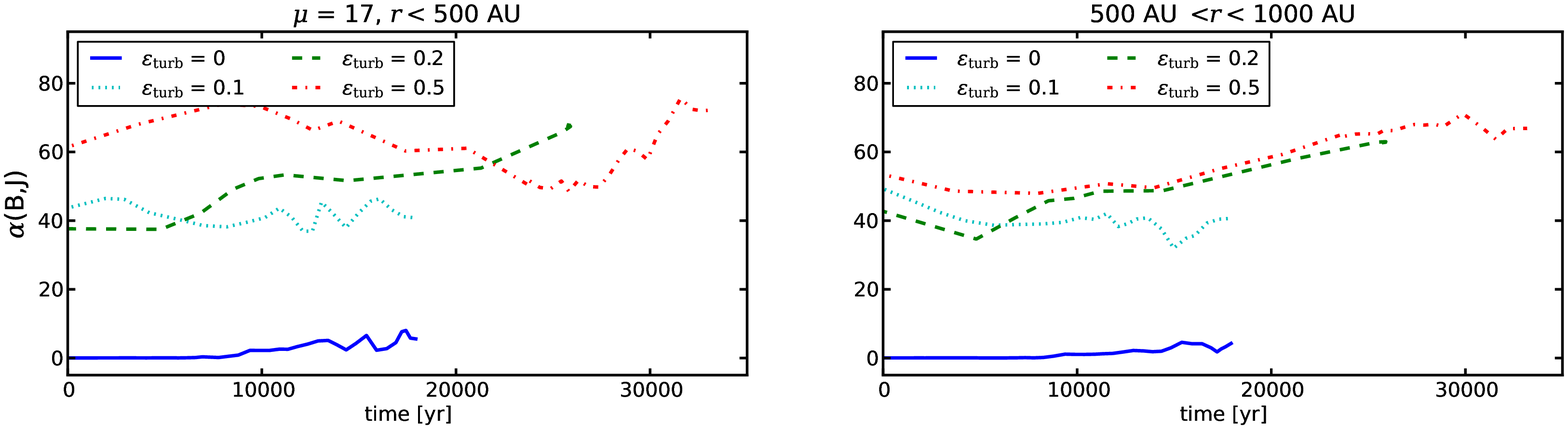}}
  \subfigure[\label{img:angle5}]{\includegraphics[width=0.89\textwidth]{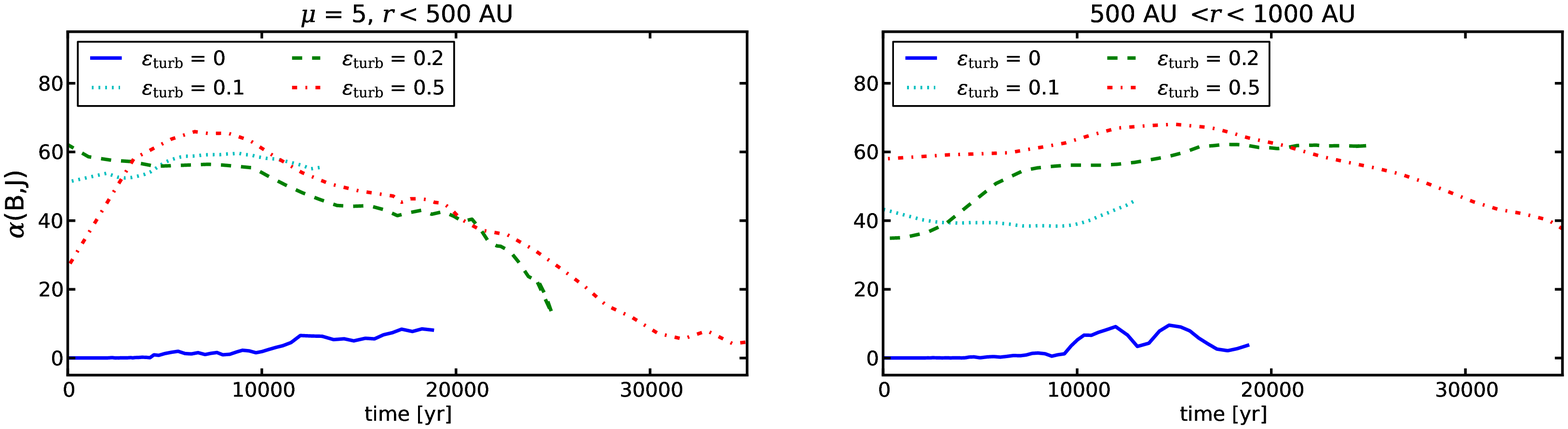}}
  \subfigure[\label{img:angle2}]{\includegraphics[width=0.89\textwidth]{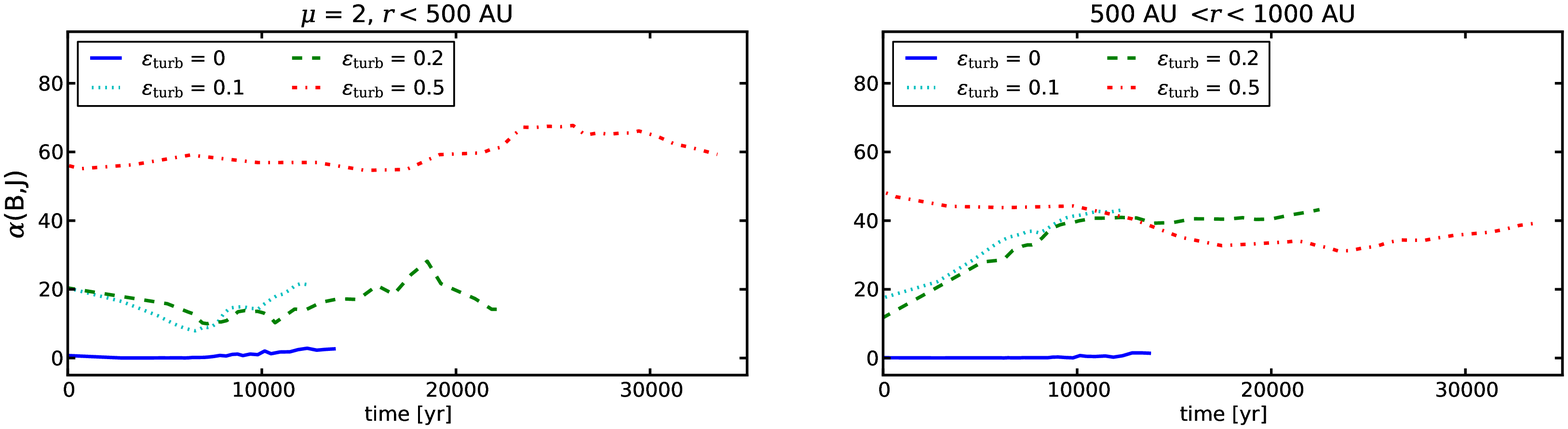}}
  \caption{Angle between the angular momentum and the magnetic field for $\mu=$ 17 (upper panels), 5 (middle panels) and 2 (lower panels), for the four levels of turbulence (\eturb{} = 0, 0.1, 0.2, and 0.5). Their respective orientation is showed in two different regions: one defined by $100<r<500$~AU (left panels) and the other by $500<r<1000$~AU (right panels).} 
\label{img:angle} 
\end{figure*}
 
Figure \ref{img:angle} shows the time evolution of the angle between the angular momentum and the magnetic field for $\mu=$ 17, 5, and 2, and \eturb{} = 0, 0.1, 0.2, and 0.5. The misalignment angle $\alpha$ is calculated for two regions: $100<R<500$ AU, and $500<R<1000$ AU. For smaller radii, the magnetic field is strongly twisted around the first core and defining a mean direction of the field is unrealistic. While the first region corresponds to the disk formation region, we also consider a larger scale since the orientation is not as clearly defined in the turbulent cases as in the laminar case.

Figure \ref{img:angle} clearly shows that without turbulence, the magnetic field and the angular momentum stay well aligned during the collapse of the first core and the evolution of the first core. On the contrary, with turbulence, the angular momentum and the magnetic field are strongly misaligned. The angle $\alpha$ ranges from 20 to 80$^{\circ}$. In addition, the mean angle changes between the inner and the outer regions, indicating that the structure of the magnetic field is complex. Again, the results for the sub- and supersonic cases are very similar. Even a subsonic turbulent velocity field is sufficient to tilt the rotation axis of the first core. 

\subsection{Magnetic braking}

To study the transport of angular momentum we estimate the magnetic braking in a cylinder defined by a radius $R=300$~AU and a height $H=300$~AU. The vertical and radial components of the integrated flux of angular momentum transported by the magnetic field are defined as 
\begin{eqnarray*}
F_z(R)=\\
\left|\int_0^{2\pi}\int_0^R\right. &  & \left.r\frac{B_{\phi}(r,\phi,\pm h(r)/2)B_z(r,\phi,\pm h(r)/2)}{4\pi}r{\rm d}r{\rm d}\phi\right|,
\end{eqnarray*}
 and 
\begin{eqnarray}
F_r(R)=\label{eq:magbrakcomp}\\
\left|\int_0^{2\pi}\int_{-h(R)/2}^{h(R)/2}\right. &  & \left.R\frac{B_{\phi}(R,\phi,z)B_r(R,\phi,z)}{4\pi}R{\rm d}z{\rm d}\phi\right|,\nonumber 
\end{eqnarray}
 where $B_i(R,\phi,z) \equiv B_i(r=R,\phi,z)$. We note that $F_z$ is the sum of the fluxes through the faces defined by $H/2$ and $-H/2$. In the following, we study the specific quantities $F_r/M$ and $F_z/M$, where $M$ is the mass enclosed in the volume of interest.

\begin{figure*}
\includegraphics[width=1\textwidth]{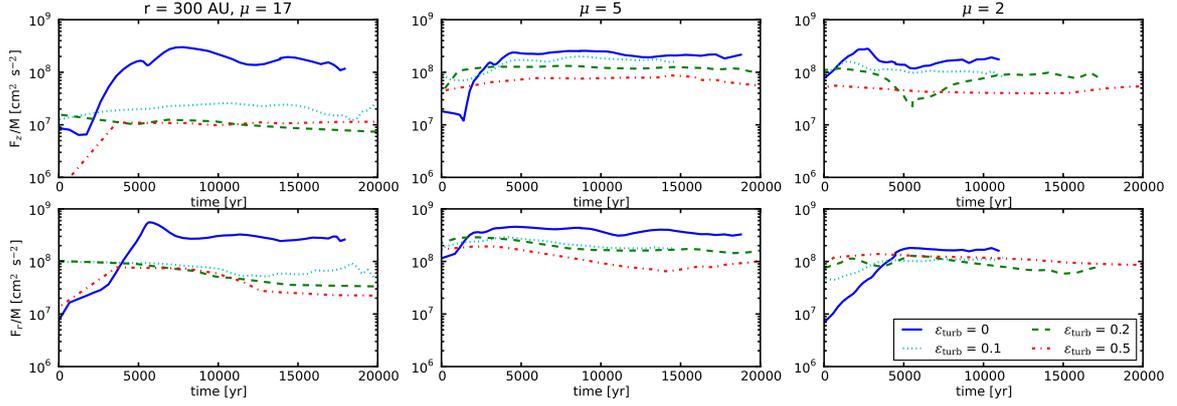}
 \caption{Evolution of the vertical and radial components of the flux of specific angular momentum transported by the magnetic field in logarithmic scale, for $\mu=17$, 5 and 2 and \eturb{} = 0, 0.1, 0.2, and 0.5. $t=0$ corresponds to the formation of the first core.}  
\label{img:magBrComp} 
\end{figure*}
 
Figure~\ref{img:magBrComp} shows the evolution of the vertical (upper panels) and radial components (lower panels) of the above-mentioned fluxes for $\mu=17$, 5 and 2, and for \eturb{} = 0, 0.1, 0.2, and 0.5.

The magnetic braking is globally less efficient in the turbulent cases than in the laminar case. For \eturb{} = 0.1, 0.2 and 0.5, both components $F_z$ and $F_r$ (upper and lower panels of Fig.~\ref{img:magBrComp}) are weaker by a factor of about 2 compared to the laminar case. In the subsonic case (\eturb{} = 0.1) the flux is stronger than in the supersonic cases (\eturb{} = 0.2 and 0.5), although weaker than in the laminar case. In general, the larger the turbulent energy, the weaker the fluxes (both components $F_z$ and $F_r$).

We also note that after a transitory phase, these fluxes become almost constant in time. However, while the generation of $B_r$ and $B_{\phi}$ due to turbulent motions is fast, and $F_z$ and $F_r$ grow rapidly, in the laminar case their increase is relatively slow ($\sim$5000 years).

To summarize, the fluxes of angular momentum transported by the magnetic field are weaker in the turbulent cases because (i) of the turbulence-induced magnetic diffusivity which reduces the magnetic flux in the central regions, (ii) of the misalignment between the rotation axis and magnetic field, which decrease the magnetic braking efficiency.

\section{Consequences for star formation}
\subsection{Disk formation}
\subsubsection{Disk definition}

We have seen in the previous section that the presence of a turbulent velocity field can considerably reduce the magnetic braking in the inner regions of the first core. We now study its consequences on disk formation by following the analysis developed in \cite{Joos12}, which is first briefly recalled.

We work in the frame of the disk, which is a cylindrical frame whose axis is parallel to the direction of the angular momentum of the first core. Since disks are expected to be reasonably axisymmetric, they are defined on concentric and superposed rings in which density, velocity, pressure and magnetic field are averaged. As discussed in \cite{Joos12} a single rotation criterion is not enough to define disks. We use instead a combination of five criteria, which are: 
\begin{enumerate}
\item the first criterion ensures that rotational support is significant: $v_{\phi}>fv_{r}$; 
\item the second criterion ensures that the disk is reasonably close to hydrostatic equilibrium: $v_{\phi}>fv_{z}$; 
\item to check whether rotational support dominates thermal support, we require: $\rho v_{\phi}^{2}>fP_{{\rm th}}$; 
\item a connectivity criterion; 
\item a density criterion is used to avoid large spiral arms: $\rho>10^{9}$~cm$^{-3}$. 
\end{enumerate}
For the first three criteria, we take a value $f=2$. 

\subsubsection{Mass of the disk}

\begin{figure}
\subfigure[\label{img:Mdisk17}]{\includegraphics[width=.5\textwidth]{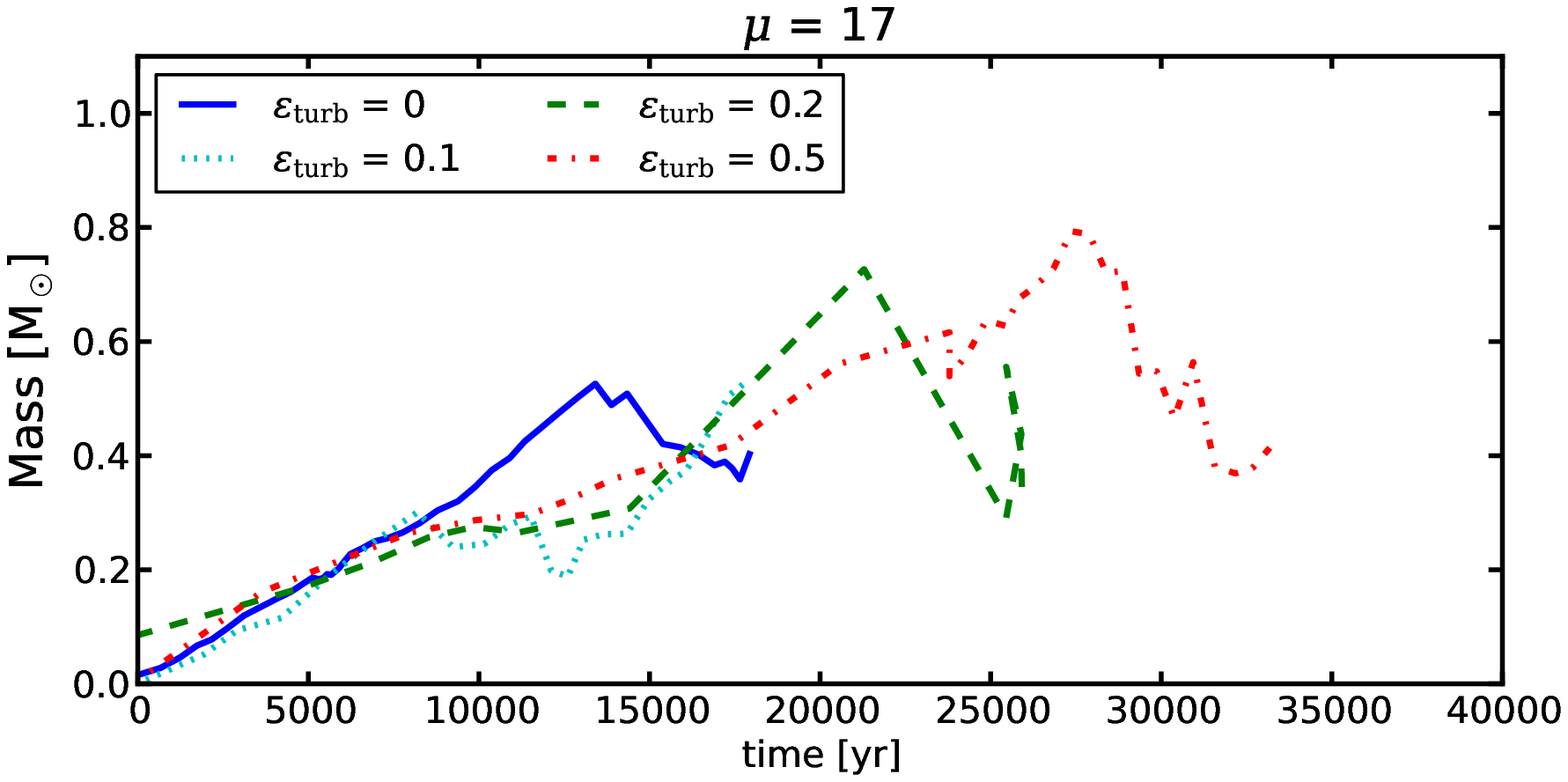}}
\subfigure[\label{img:Mdisk5}]{\includegraphics[width=.5\textwidth]{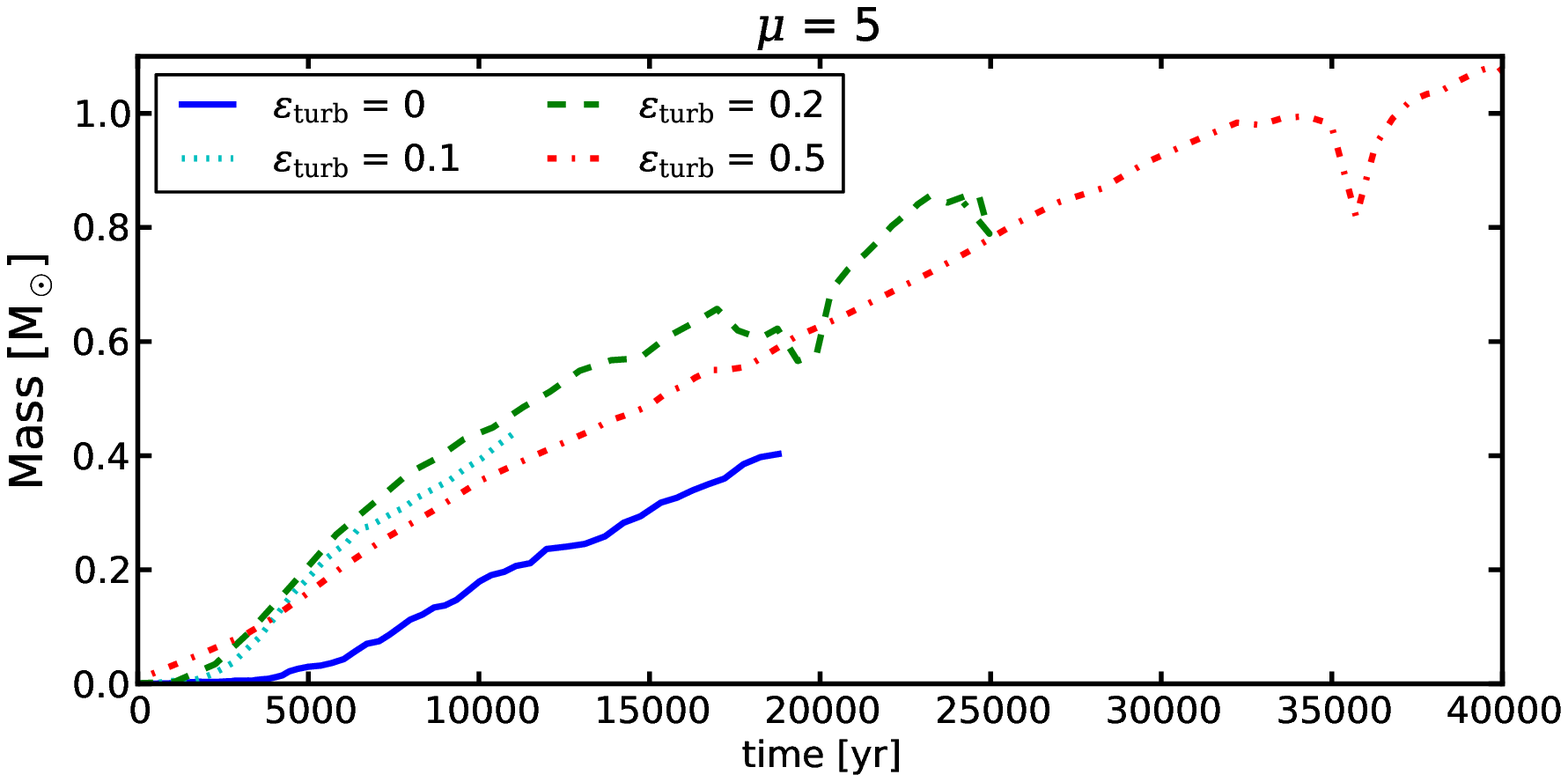}}
\subfigure[\label{img:Mdisk2}]{\includegraphics[width=.5\textwidth]{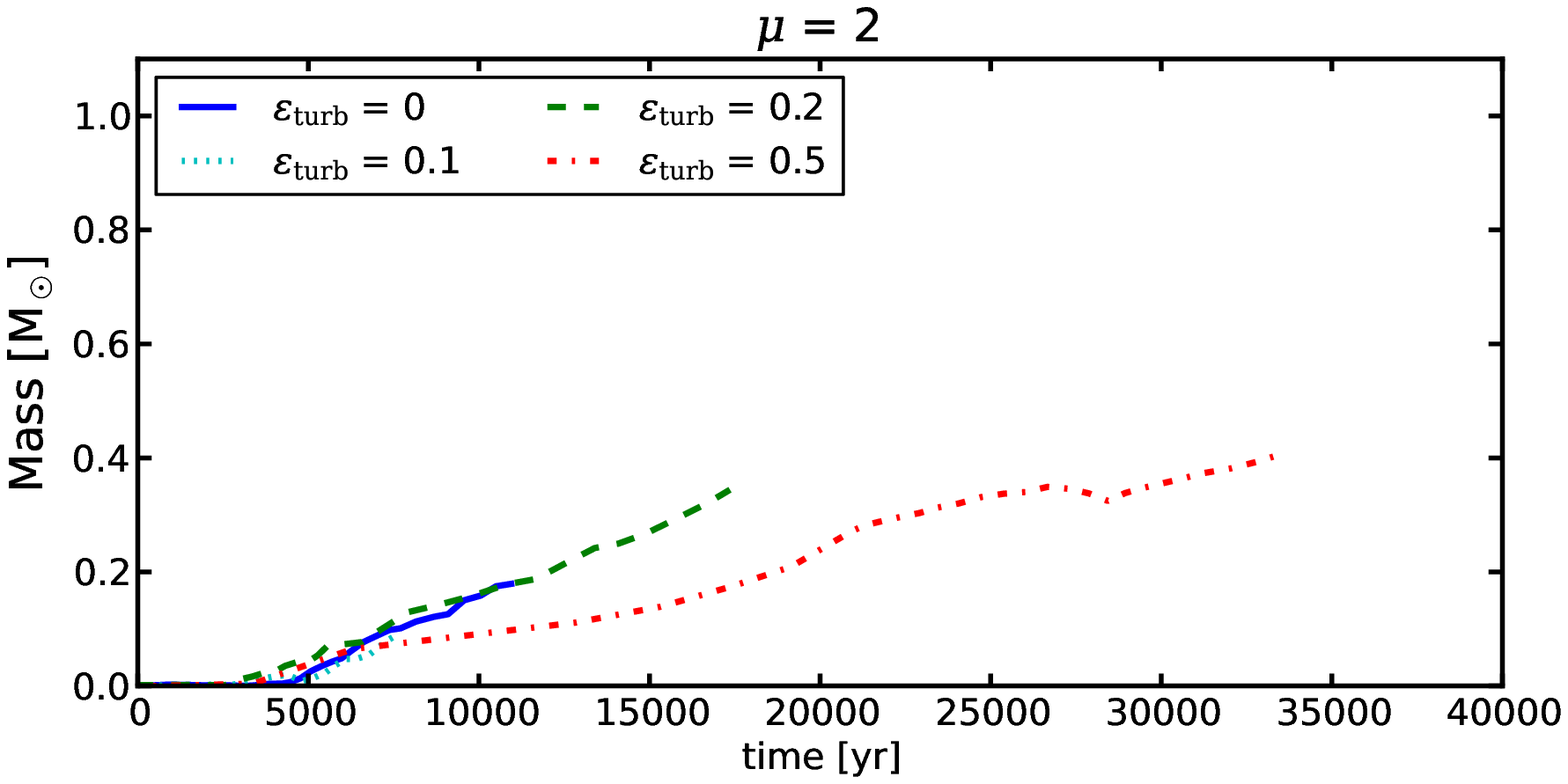}}
\caption{Mass of the disk as a function of time for $\mu = 17$ (Fig. \ref{img:Mdisk17}), $\mu = 5$ (Fig. \ref{img:Mdisk5}) and $\mu = 2$ (Fig. \ref{img:Mdisk2}).}
\label{img:Mdisk}
\end{figure}

The mass of the disk as a function of time, for different magnetizations and turbulence levels, is presented in Fig.~\ref{img:Mdisk}. For each case $t=0$ corresponds to the time when the first core forms. Because of the weak magnetic braking, disks always form for $\mu=17$. A result that is not changed significantly by turbulence. The observed rapid decrease in the masses of the disks is caused by fragmentation, which is discussed in the next section. For $\mu=5$ massive disks can form in the presence of turbulence, their masses growing up to 1.0~$M_{\odot}$, which represents more than 20\% of the prestellar core's mass. This is in contrast with the laminar (and aligned) case, where only relatively smaller disks are able to form. For the more magnetized case, $\mu=2$, massive disks do not form at all, with or without turbulence, instead disks with masses in the range of about 0.3 to 0.4~$M_{\odot}$ can form. We note that because of numerical convergence issues, these masses are probably overestimated by a factor of at least 2 (see appendix~\ref{sec:conv}).

In the laminar case, for the intermediate magnetization ($\mu=5$), there are two regimes: before $\sim$5000~yr the mass of the disk is negligible, and after $\sim$5000~yr it starts to grow with a slope comparable to the turbulent cases (Fig.~\ref{img:Mdisk5}). We believe that it is due to a symmetry breaking occurring at this stage, due to the interchange instability discussed in \cite{Krasnopolsky12}. This symmetry breaking generates turbulent motions in the vicinity of the first core and causes a transport of magnetic flux that can be seen in the first panel of Fig.~\ref{img:diffmu05} where two regimes can also be seen before and after $\sim$3000~yr. This magnetic flux diffusion induces a decrease of the magnetic braking efficiency; more angular momentum is then available to build disk. This effect probably depends on the non-ideal MHD processes taken into account (and on the numerical methods used when these processes are not explicitly treated), and would require further investigations. This is consistent with our previous claim that the aligned configuration is peculiar.

These results show that due to the reduced magnetic braking, in the low and intermediate magnetization cases there is a trend to form bigger disks when $\eturb\neq0$. At the same time, it is still clear that for increasing magnetic field strength, and thus magnetic braking, disks tend to be smaller. More precisely, for $\mu=2$, disks with masses greater than 10\% of the mass of the prestellar core can hardly form, at the time simulated. Importantly, in the intermediate magnetization case $\mu = 5$, the mass of the disks in the subsonic case (\eturb{} = 0.1) tends to be only slightly smaller than in the supersonic cases. Nevertheless the mass of the disks is significantly larger than that of the disks built in a non-turbulent environment. Therefore, even a low level of turbulence is sufficient to form massive disks when $\mu\gtrsim5$.

These results depend also on the realisation of the turbulent velocity field, as discussed in section~\ref{sec:seed}, and show a large variance with the realisations. However, the main trends do not change. In most of the cases studied, turbulence favours the formation of more massive disks.

\subsection{Fragmentation}

\begin{figure}[h]
  \subfigure[\label{img:massfrag510}]{\includegraphics[width=.5\textwidth]{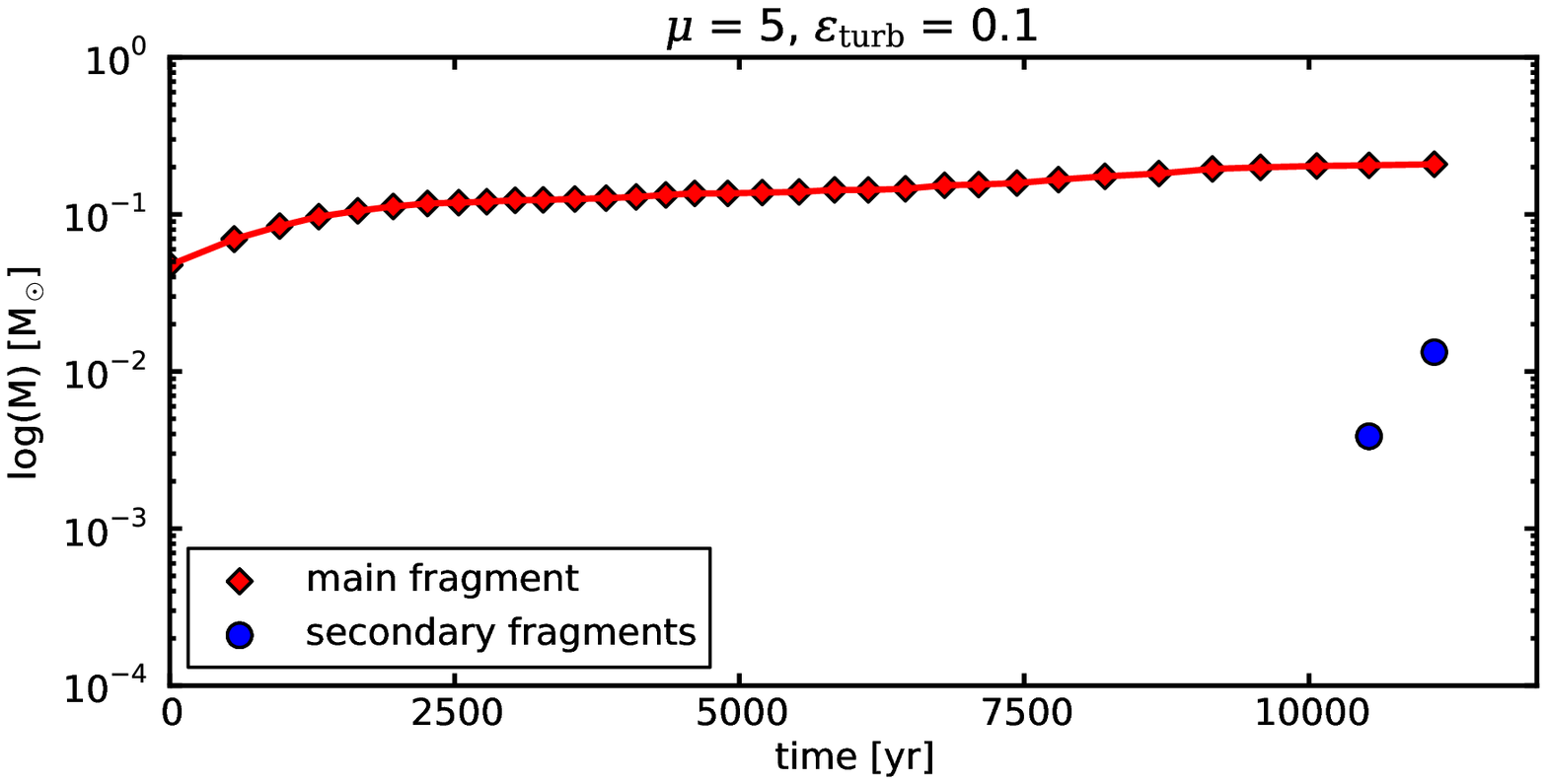}}
  \subfigure[\label{img:massfrag520}]{\includegraphics[width=.5\textwidth]{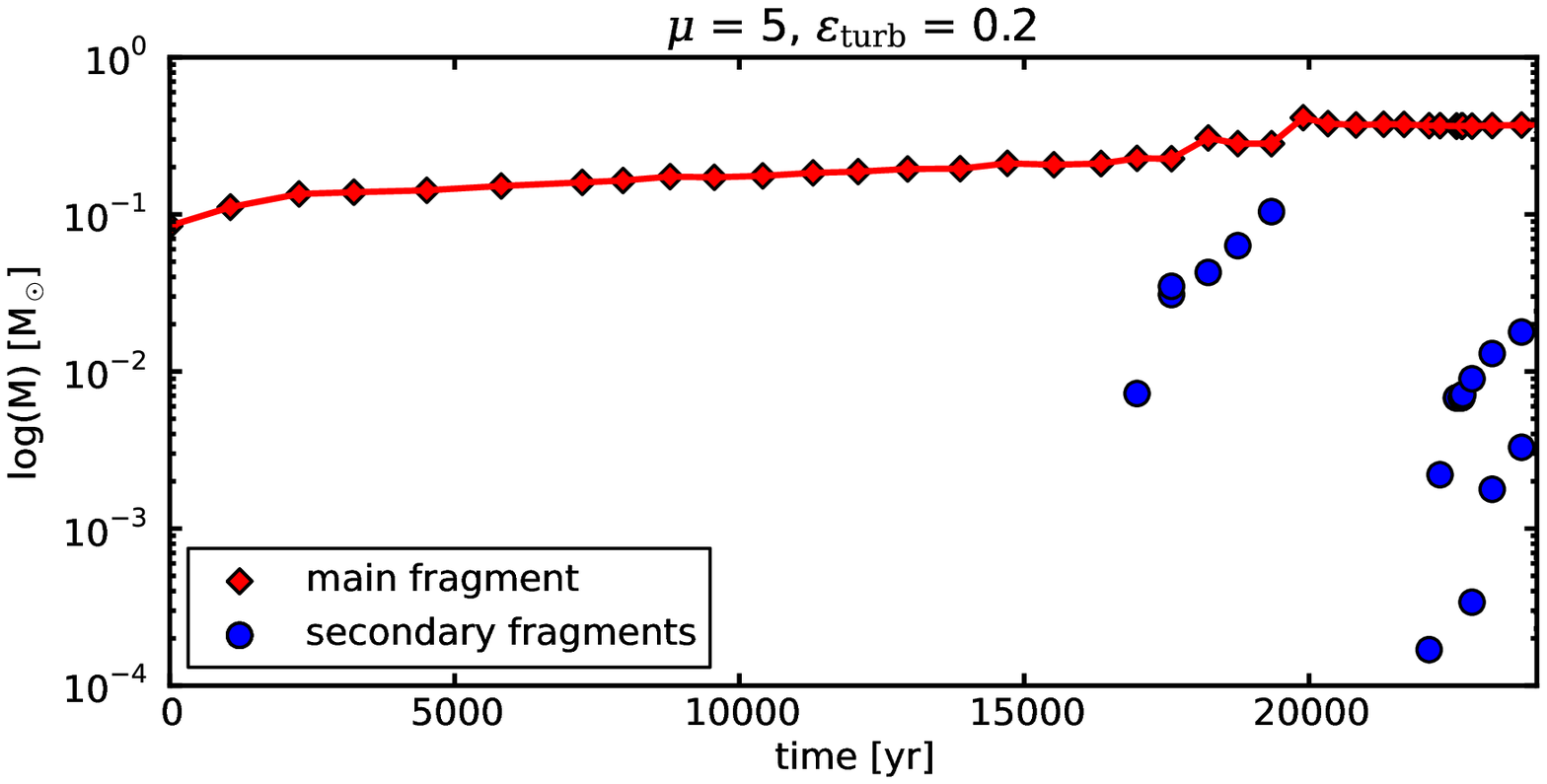}}
  \subfigure[\label{img:massfrag550}]{\includegraphics[width=.5\textwidth]{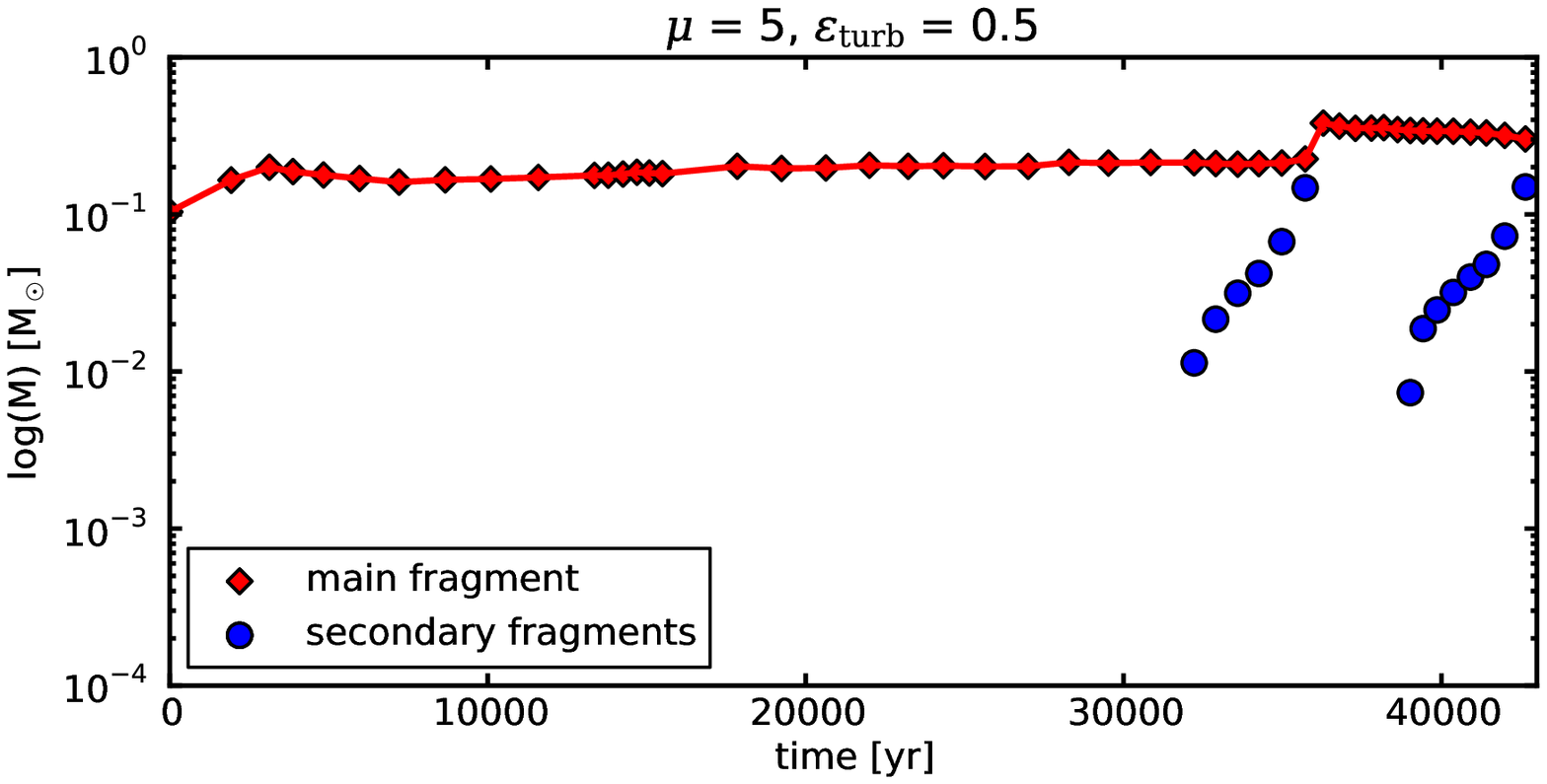}}
  \caption{Mass of the fragments for $\mu = 5$. The central, most massive first core is denoted by the red diamonds and line.}
  \label{img:massfrag5}
\end{figure}

\begin{figure}[h!]
  \subfigure[\label{img:frag170}]{\includegraphics[width=.47\textwidth]{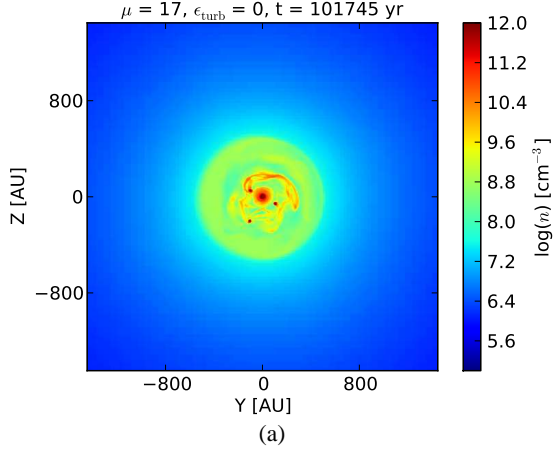}}
  \subfigure[\label{img:frag1720}]{\includegraphics[width=.47\textwidth]{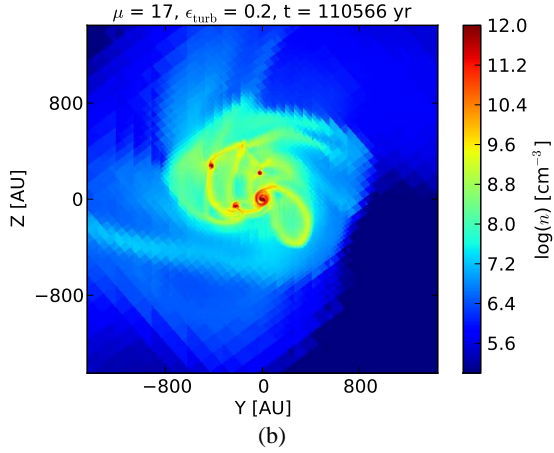}}
  \caption{Slice in density in the equatorial plane for $\mu = 17$, without turbulence (upper panel), and for \eturb{} = 0.2 (lower panel).}
  \label{img:fragslice1}
\end{figure}

\begin{figure}
  \subfigure[\label{img:frag1750}]{\includegraphics[width=.47\textwidth]{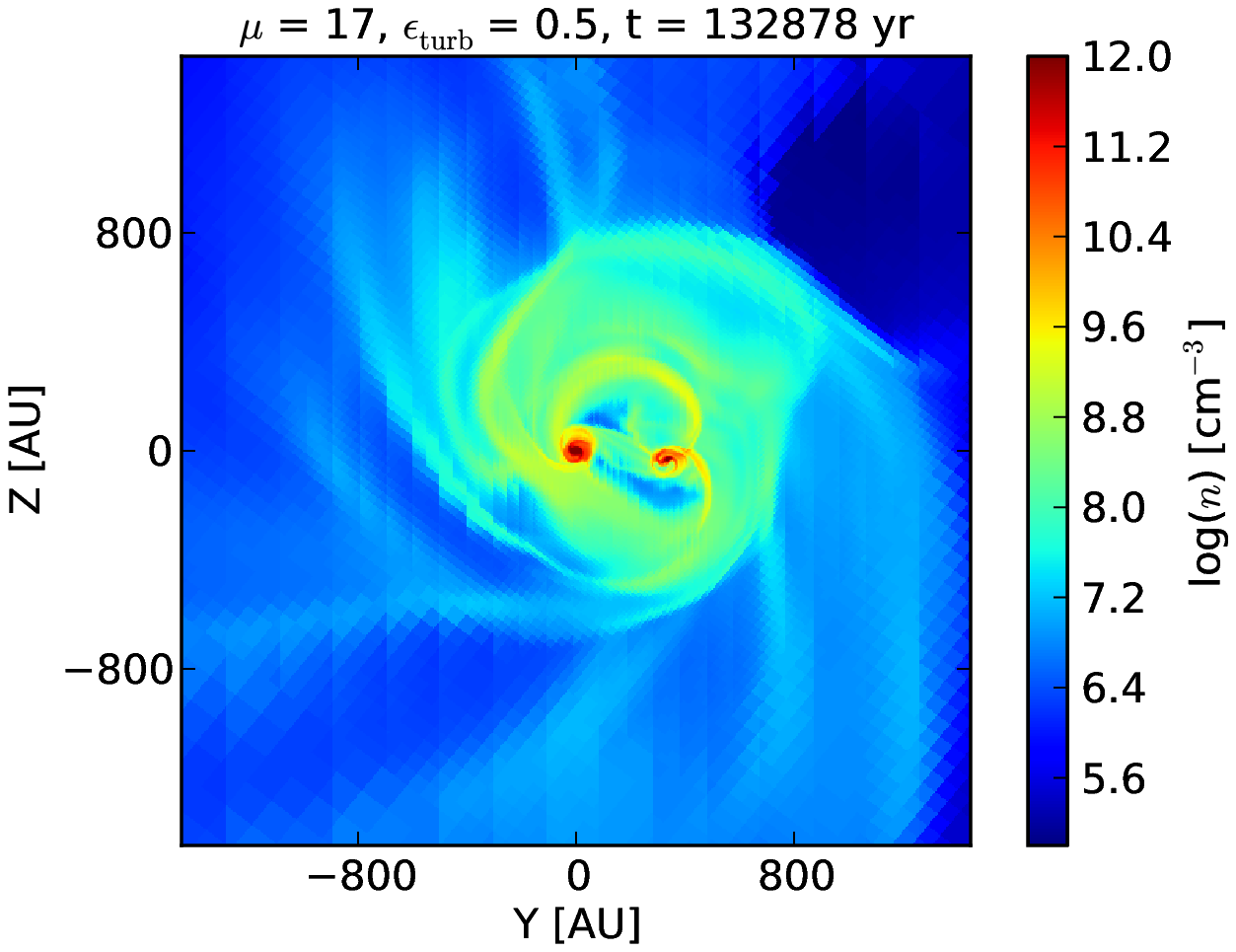}}
  \caption{Same as Fig.~\ref{img:fragslice1} for \eturb{} = 0.5.}
  \label{img:fragslice2}
\end{figure}

\begin{figure}
  \subfigure[\label{img:massfrag170}]{\includegraphics[width=.5\textwidth]{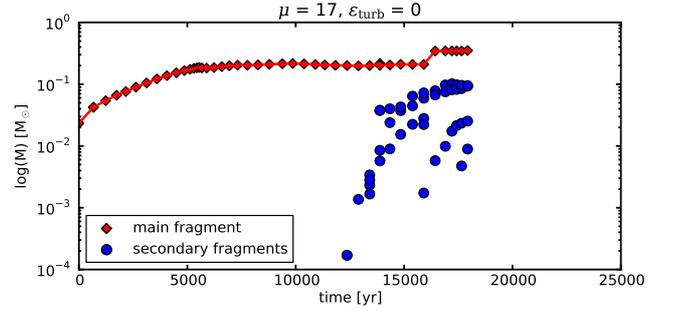}}
  \subfigure[\label{img:massfrag1710}]{\includegraphics[width=.5\textwidth]{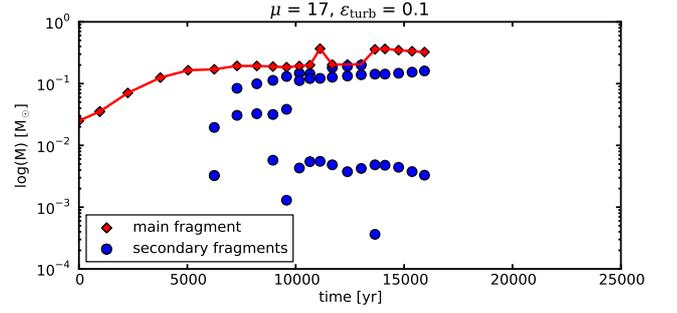}}
  \subfigure[\label{img:massfrag1720}]{\includegraphics[width=.5\textwidth]{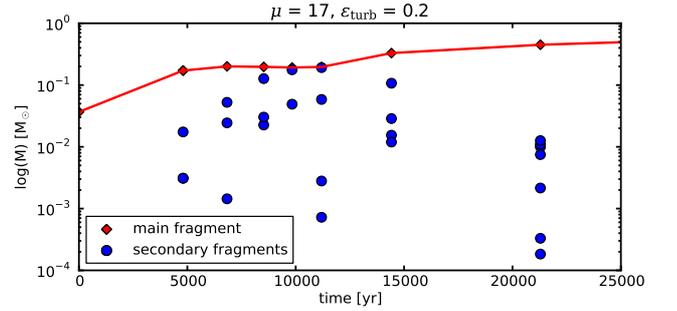}}
  \subfigure[\label{img:massfrag1750}]{\includegraphics[width=.5\textwidth]{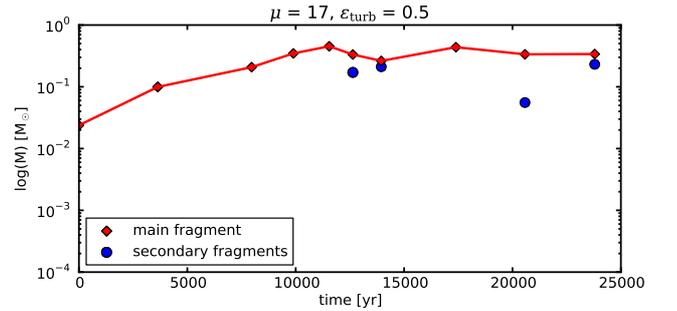}}
  \caption{Mass of the fragments for $\mu = 17$.}
  \label{img:massfrag17}
\end{figure}

Massive disks are prone to fragmenting, and to identify and study the fragments we use a friend-of-friend algorithm. In this method, a fragment is defined by starting from a pre-defined over-density, and then finding all cells above a certain density threshold which are spatially connected. Contrary to the sink particles technique, it is harder with this method to accurately follow the fragments, and they can also potentially merge. However this technique gives sufficiently accurate information on the fragmentation process occurring in our simulations. We note that to ensure the validity of those results a treatment of the second collapse, as well as radiative transfer calculations, would be needed.

As observed previously by \cite{Hennebelle08b} and \cite{Joos12} for 1~$M_{\odot}$ prestellar cores, we find that for $\mu=2$ no fragmentation occurs. The disk is too small and remains stable. Moreover, the magnetic field, and particularly its toroidal component, has a stabilizing effect that efficiently prevents fragmentation. On the other hand, the initial turbulent velocity field is too weak to provide density perturbations of sufficient amplitude that could trigger fragmentation \citep{Hennebelle08b}.

For $\mu=5$, fragmentation takes place when the disks are massive enough. This occurs at late time (approximately 20 000~yr after the formation of the first core) when the mass of the disks reaches about 0.7~$M_{\odot}$. Figure~\ref{img:massfrag5} shows, for \eturb{} = 01, 0.2, and 0.5, the evolution of the mass of the fragments: the red diamonds (and the red line) correspond to the first core itself whereas the blue dots correspond to the disk fragments. One dot corresponds to one fragment at a given time. Fragments can merge and fragment again, and up to four fragments of about $10^{-3}M_{\odot}$ form. The most massive correspond to the central first core which forms initially with a mass of about few tenths of a solar mass. We note that fragmentation occurs for these 5~$M_{\odot}$ prestellar cores whereas it was not the case for the 1~$M_{\odot}$ prestellar cores studied in \cite{Joos12}. Although the number of Jean's masses is initially the same, as the collapse occurs, gas becomes denser and temperature becomes higher in the 1~$M_{\odot}$ prestellar cores than in the 5~$M_{\odot}$ prestellar cores, simply because of the barotropic equation of state. Therefore, 5~$M_{\odot}$ prestellar cores have more Jeans' masses than 1~$M_{\odot}$ prestellar cores.

As shown by Fig.~\ref{img:fragslice1} and \ref{img:fragslice2} the fragmentation of disks also occurs for $\mu=17$. The masses of the fragments for $\mu=17$ are shown in Fig.~\ref{img:massfrag17}. Typical masses are a few tenth of a solar mass for the bigger ones, to $10^{-3}M_{\odot}$ for the smaller ones. In the turbulent cases ($\eturb\neq0$), the mass of the fragments can be comparable to the mass of the central object, whereas without turbulence, their mass is $\lesssim$30~\% of the mass of the first core. There are up to four fragments formed in the laminar and subsonic cases, up to eight fragments for \eturb{} = 0.2 and two fragments for \eturb{} = 0.5. For \eturb{} = 0.5 or higher, fragmentation is quenched because of the increased turbulent support.

\subsection{Outflows}  \label{sec:out}

\begin{figure}[h!]
  \subfigure[\label{img:out51}]{\includegraphics[width=.5\textwidth]{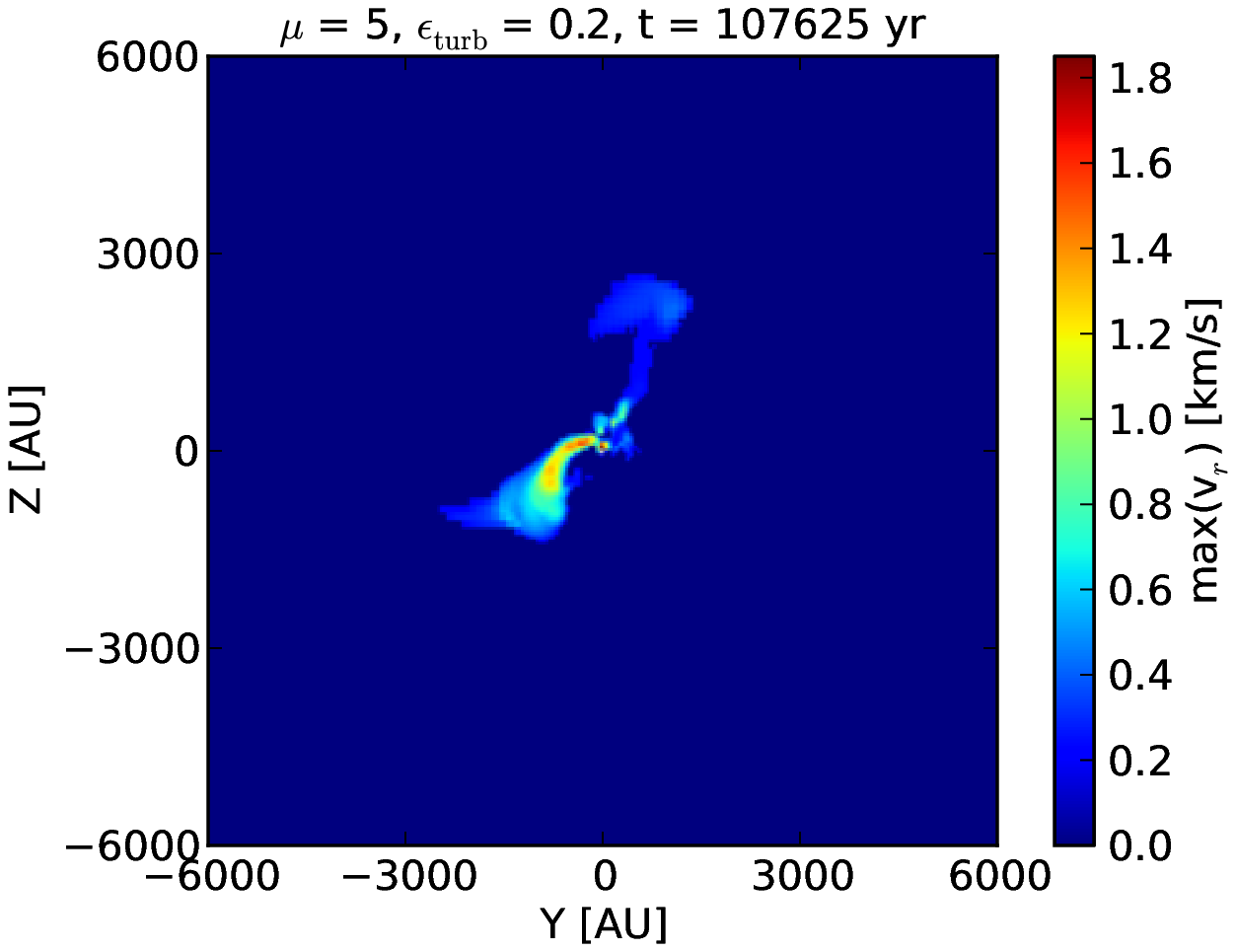}}
  \subfigure[\label{img:out52}]{\includegraphics[width=.5\textwidth]{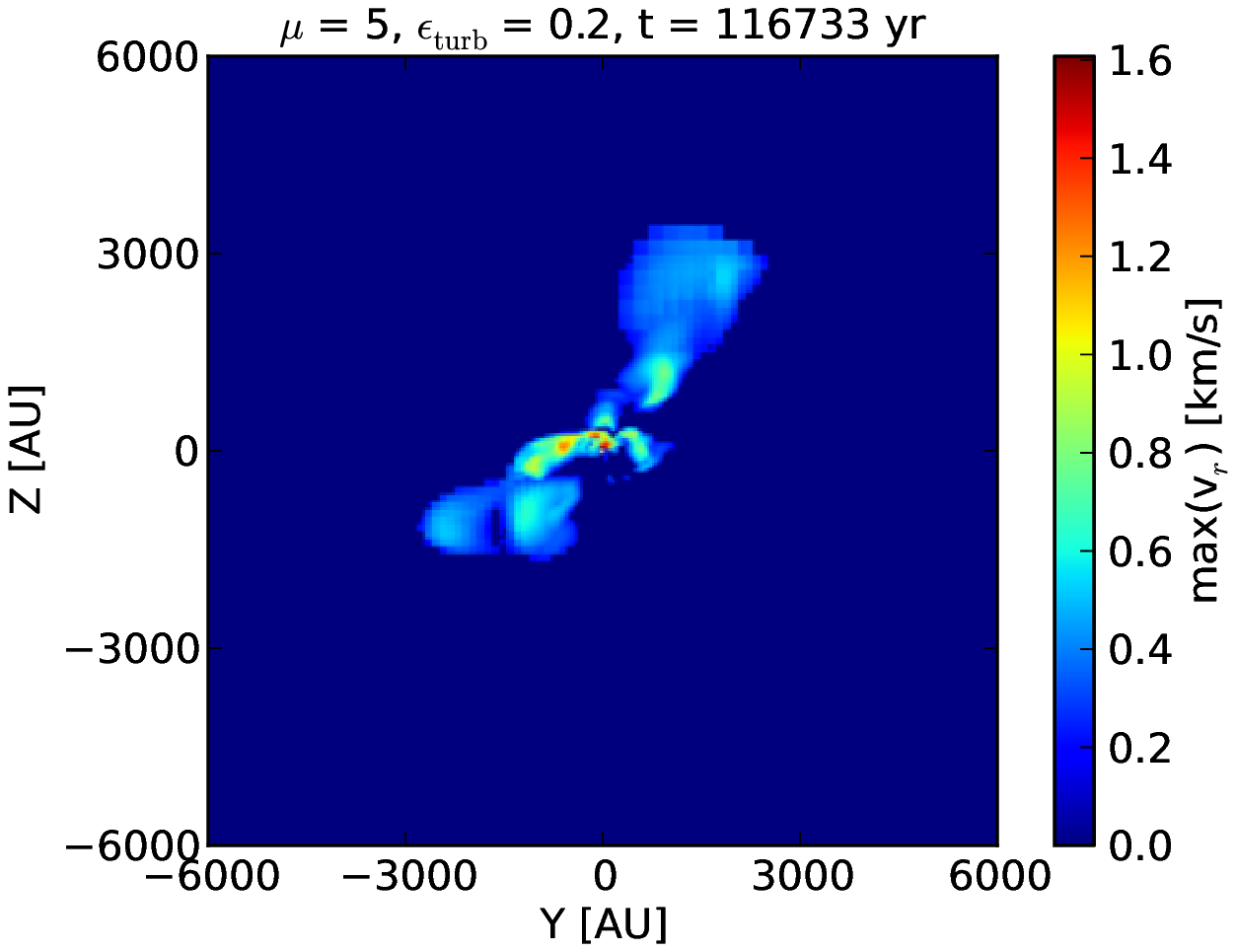}}
  \caption{Maximum of the projected radial velocity for $\mu = 5$, for \eturb{} = 0.2.}
  \label{img:out5}
\end{figure}

\begin{figure}[h!]
  \subfigure[\label{img:out21}]{\includegraphics[width=.5\textwidth]{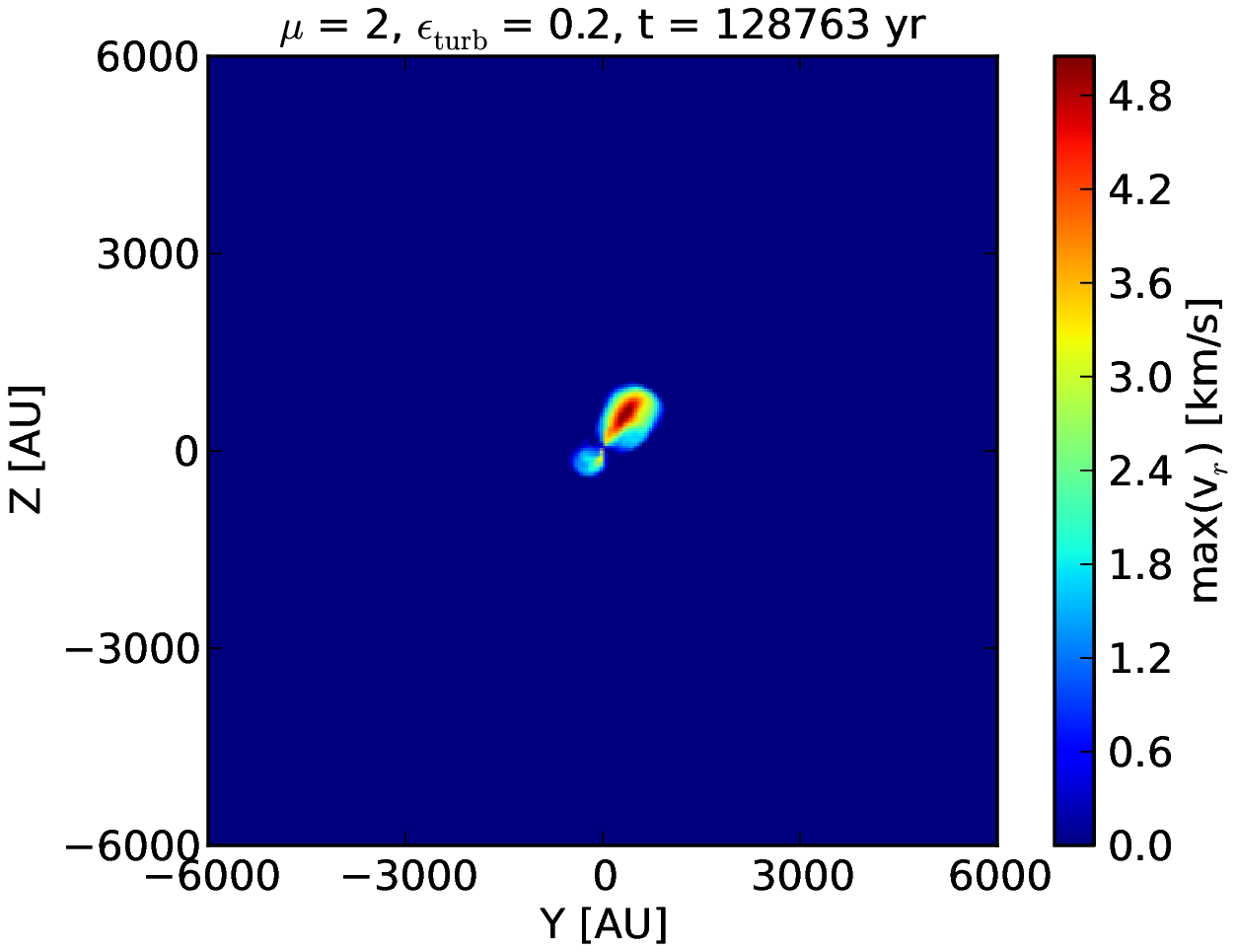}}
  \subfigure[\label{img:out22}]{\includegraphics[width=.5\textwidth]{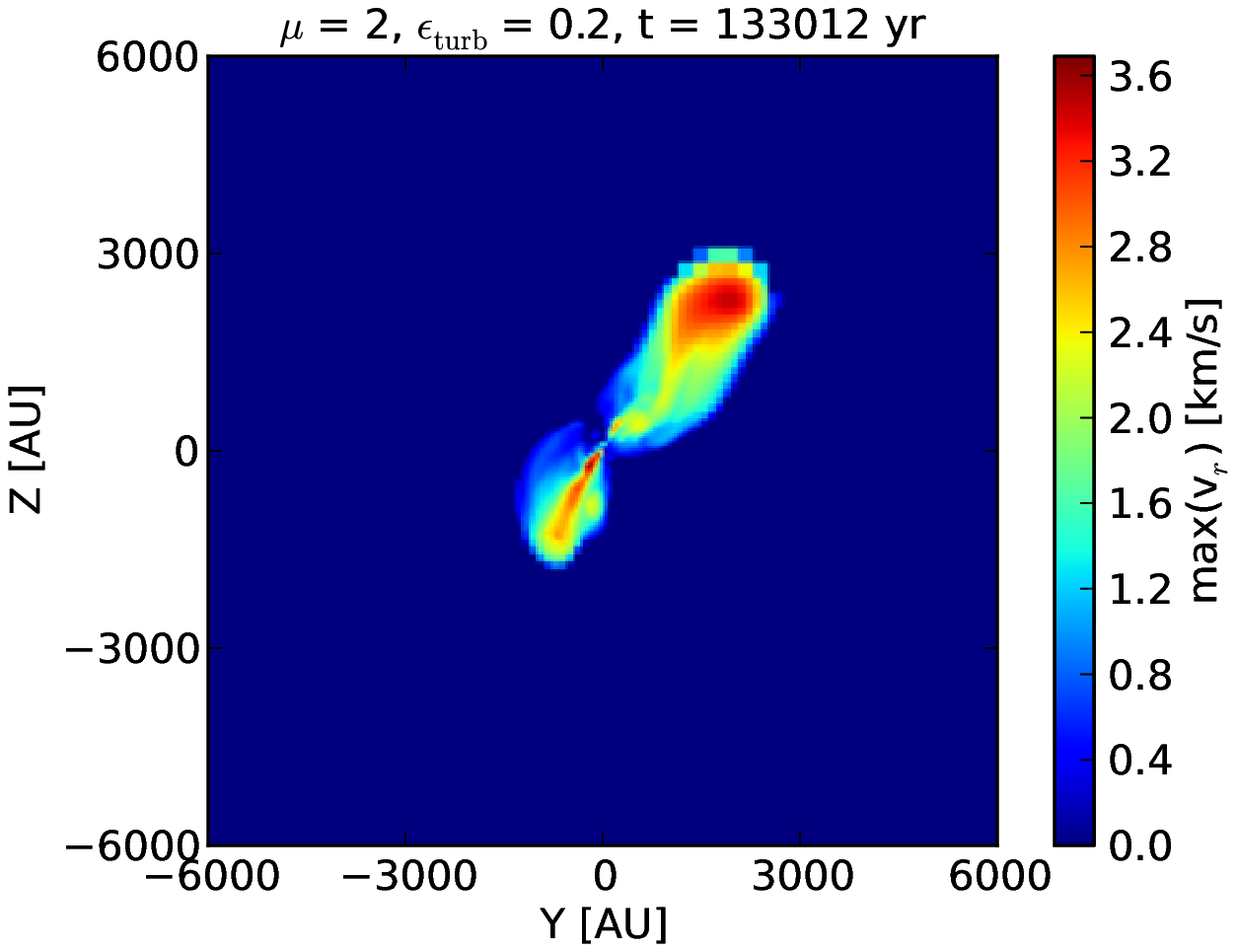}}
  \caption{Maximum of the projected radial velocity for $\mu = 2$, for \eturb{} = 0.2.}
  \label{img:out2}
\end{figure}

\begin{figure}[h!]
\subfigure[\label{img:Mout17}]{\includegraphics[width=.5\textwidth]{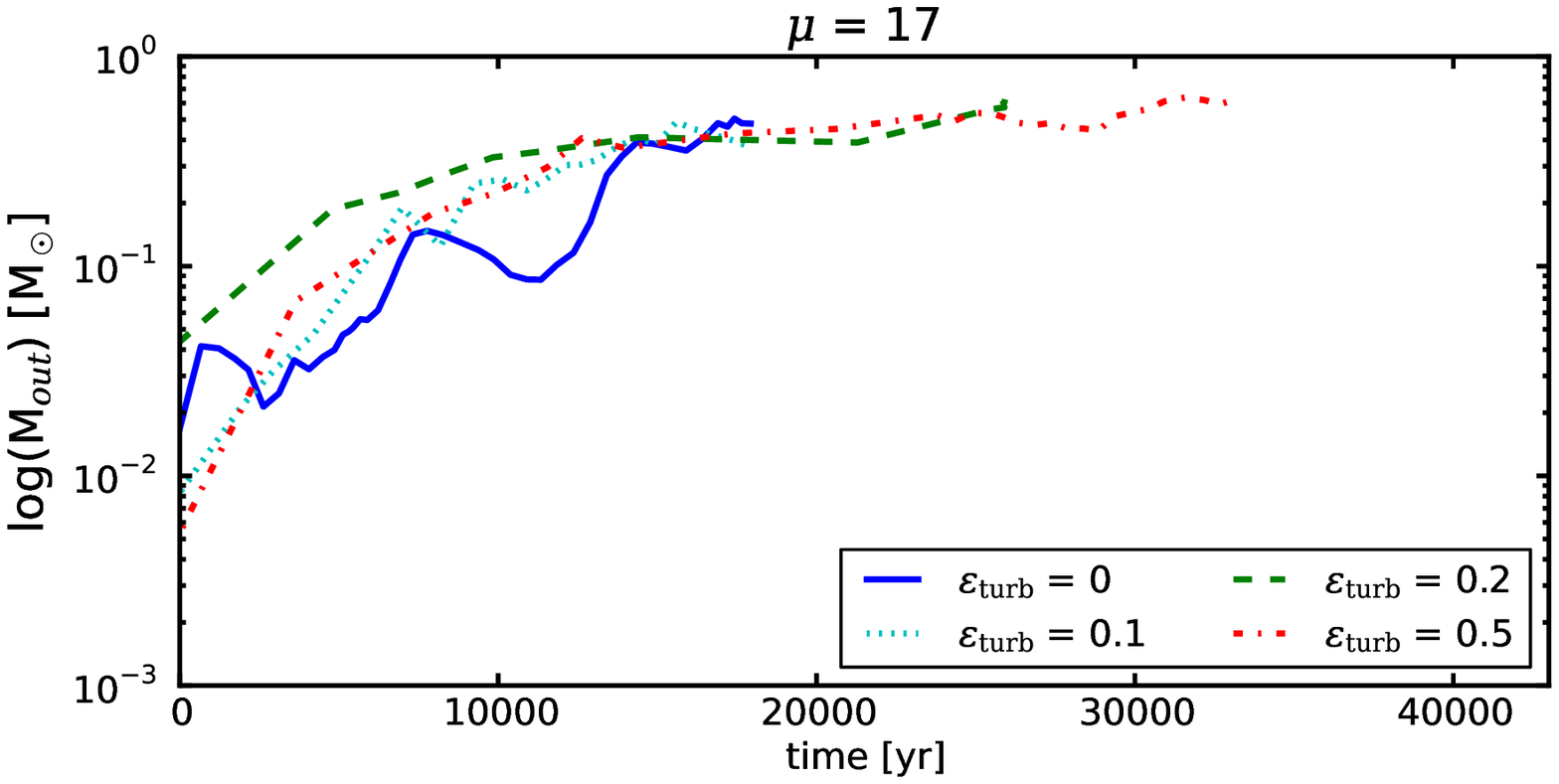}}
\subfigure[\label{img:Mout5}]{\includegraphics[width=.5\textwidth]{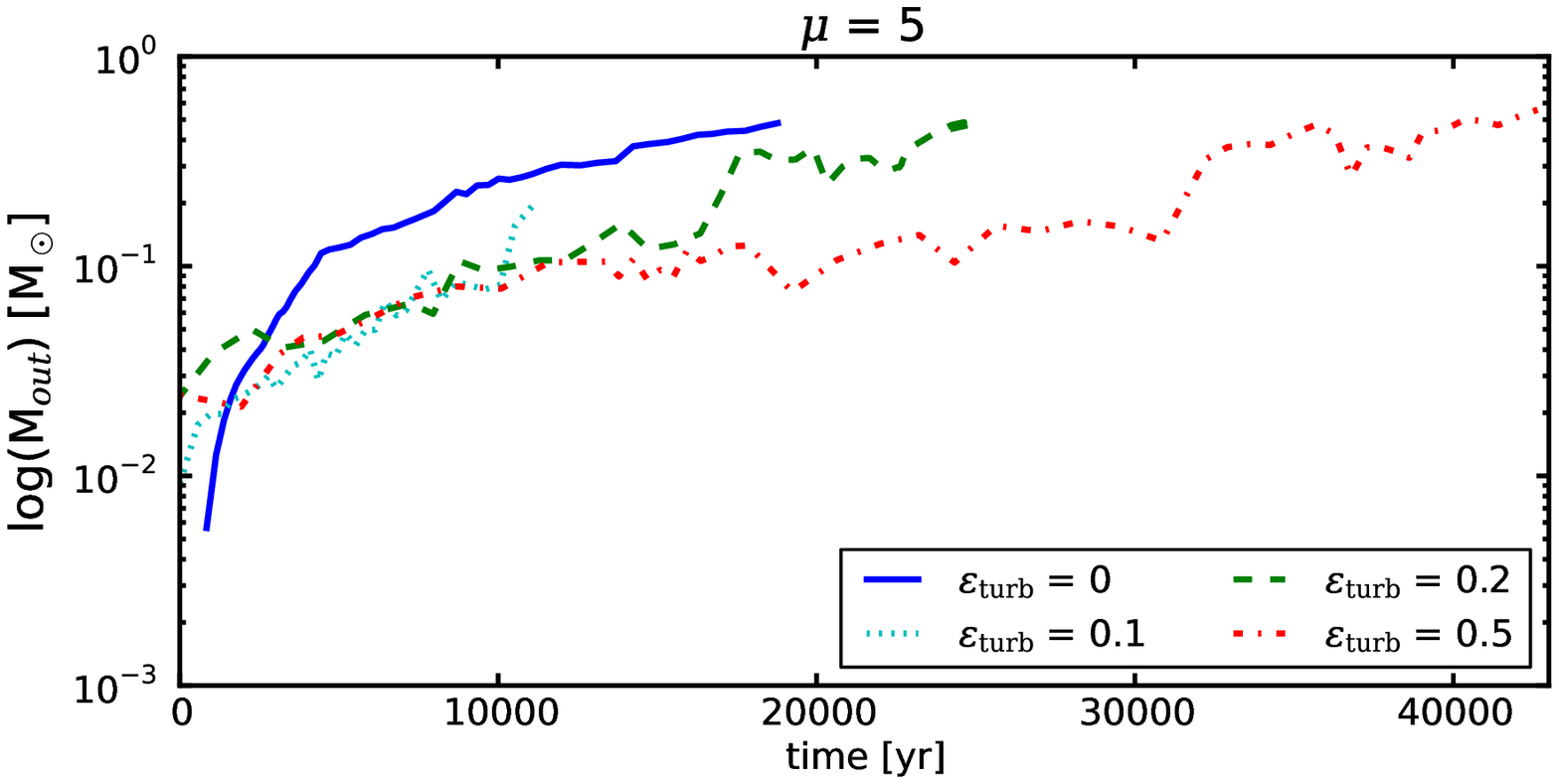}}
\subfigure[\label{img:Mout2}]{\includegraphics[width=.5\textwidth]{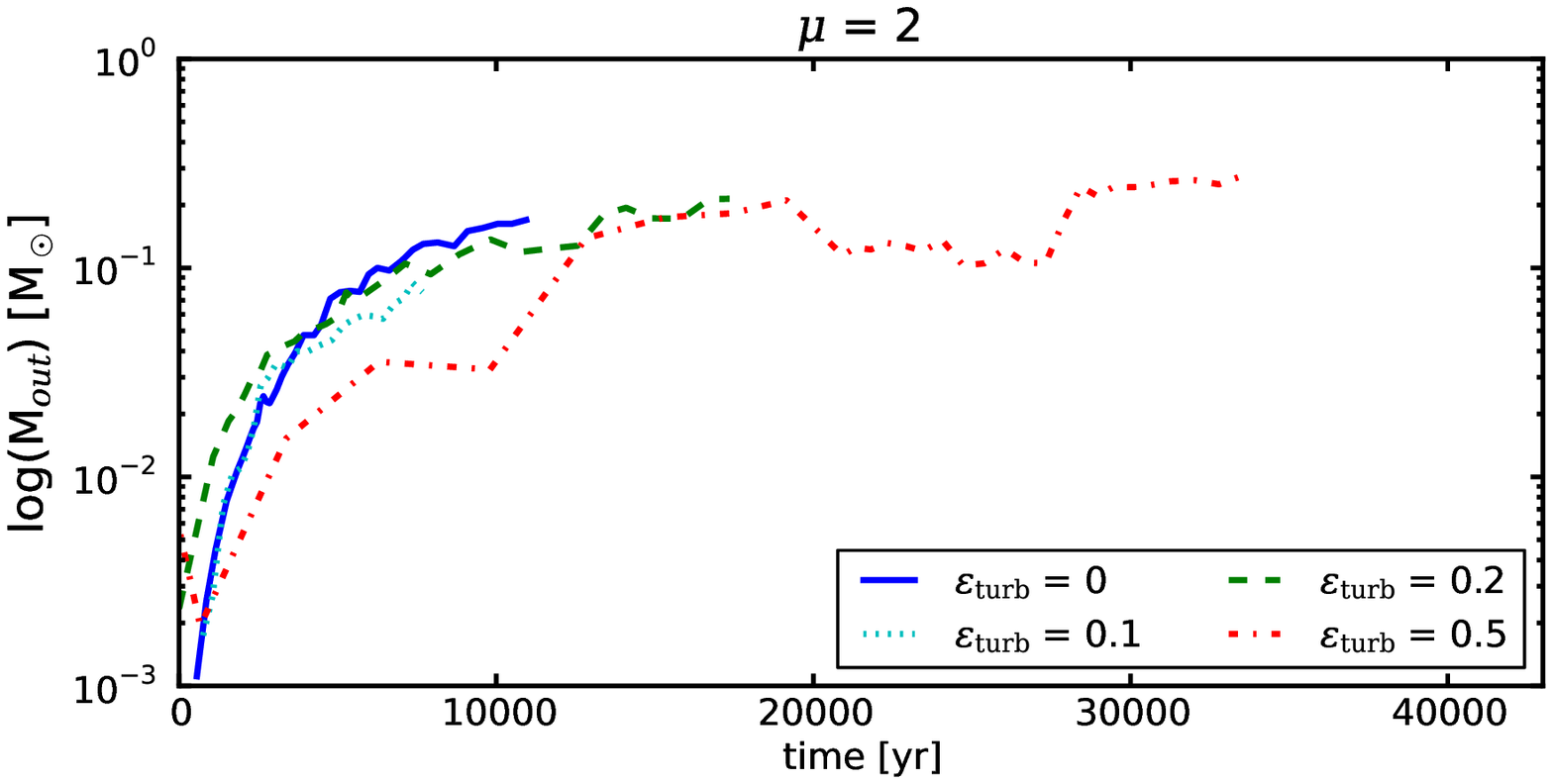}}
\caption{Total mass in the outflows for $\mu = 17, \, 5$, and 2, without turbulence, with \eturb{} = 0.1, 0.2, and 0.5.}
\label{img:Mout}
\end{figure}

\begin{figure}[h!]
\subfigure[\label{img:MoutvsMcore17}]{\includegraphics[width=.5\textwidth]{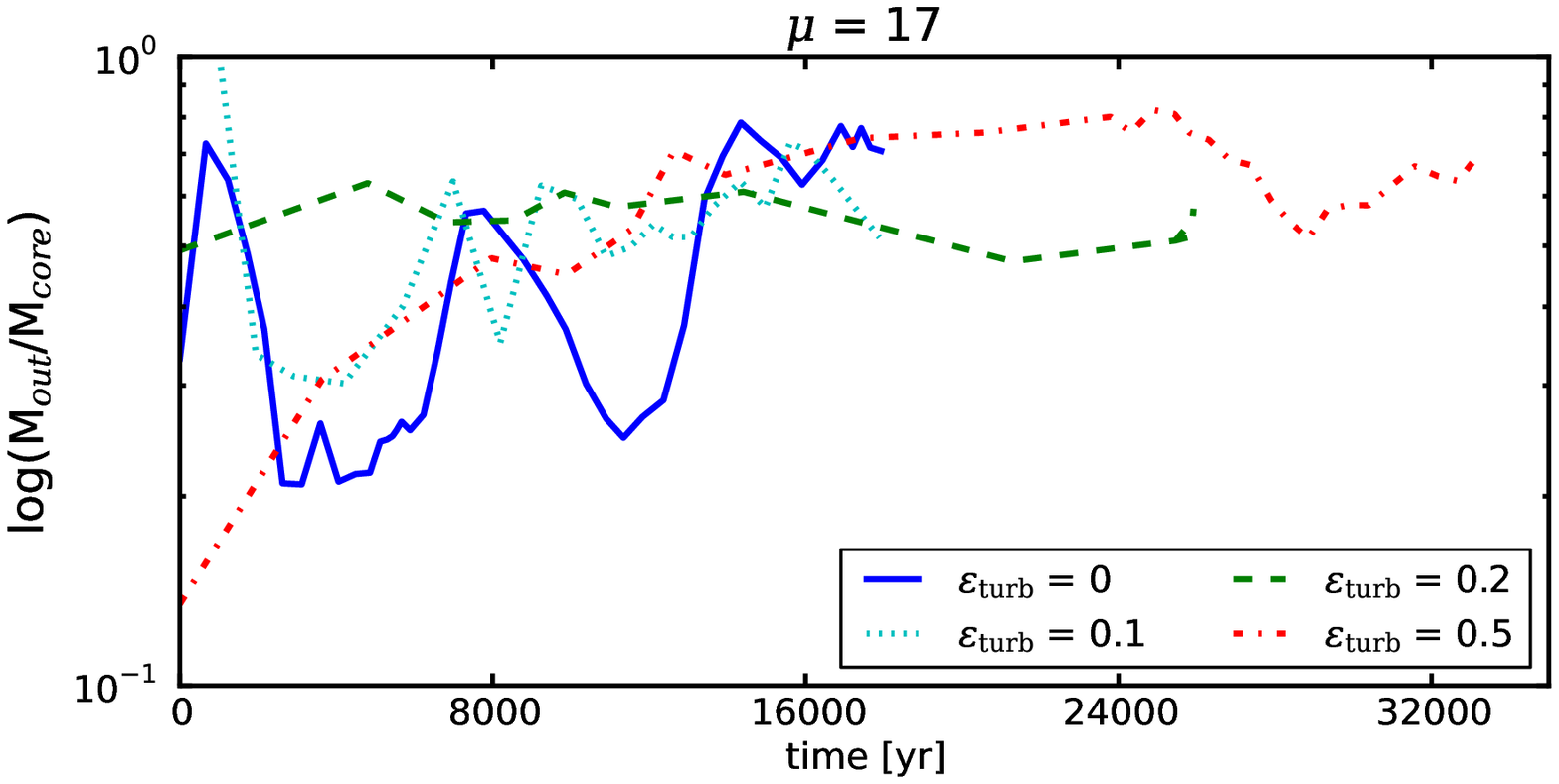}}
\subfigure[\label{img:MoutvsMcore5}]{\includegraphics[width=.5\textwidth]{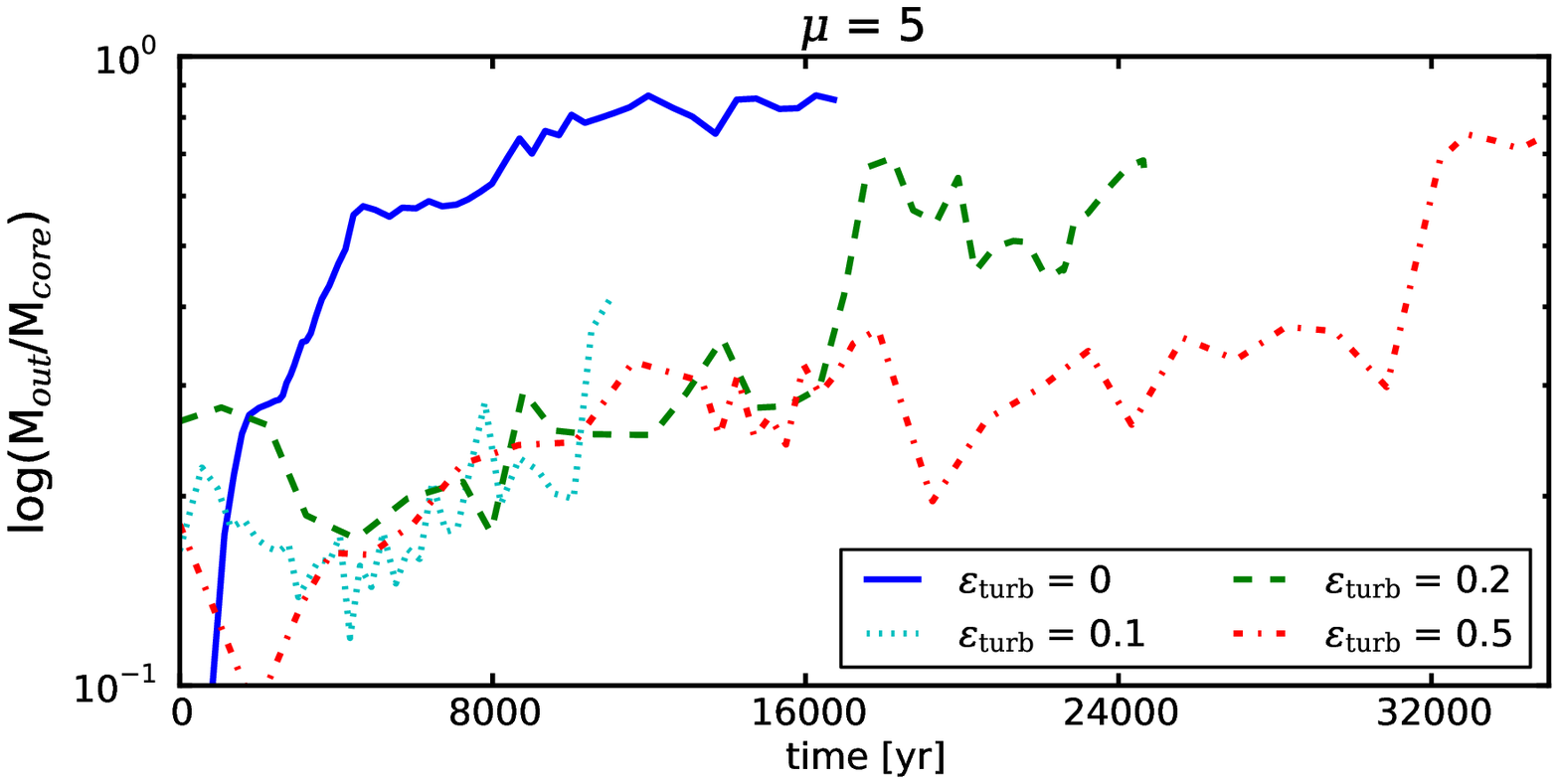}}
\subfigure[\label{img:MoutvsMcore2}]{\includegraphics[width=.5\textwidth]{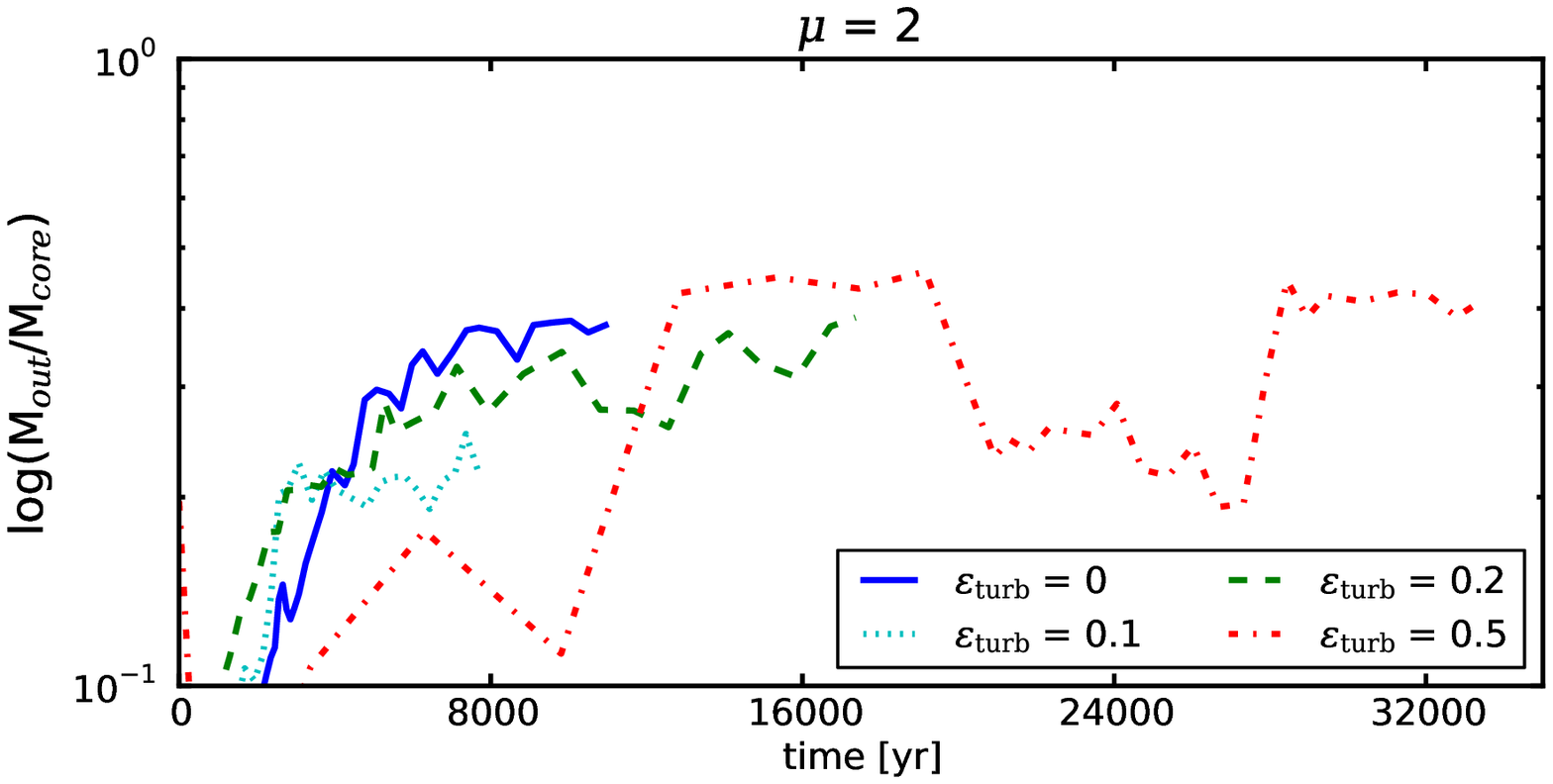}}
\caption{Ratio of the total mass in the outflows and the mass of the first core for $\mu = 17$, 5, and 2, without turbulence and with \eturb{} = 0.1, 0.2, and 0.5.}
\label{img:MoutvsMcore}
\end{figure}
 
Outflows and jets are one of the most important tracers of star formation, being associated with all phases of protostellar evolution. They are thought to be magneto-centrifugally launched \citep{Blandford82,Pudritz83,Pelletier92,Ferreira97,Uchida85}, and observations seem to confirm that they play an important role in the efficient transport of angular momentum \citep{Bacciotti02}. The early formation of outflows during the collapse of prestellar dense cores was investigated in both 2D and 3D MHD simulations \citep{Mellon08,Hennebelle08a,Ciardi10,Banerjee06, Machida08}. The formation of outflows in a turbulent medium was discussed by \cite{Hennebelle11}, however they modelled much more massive prestellar cores (100~$M_{\odot}$) than treated here.

We define outflows as the gas with ${\bf v}\cdot{\bf e}_{r}>0.2$~km s$^{-1}$, with ${\bf e}_{r}$ the radial unit vector in spherical coordinates (the center of the spherical frame being the centre of the first core itself). As a tracer of the outflow structure, Figures~\ref{img:out5} and \ref{img:out2} display the maximum of the projected radial velocity in the $yz-$plane. Despite the complexity of the magnetic field, outflows are launched in all simulations, even in the strongly turbulent cases, and reach spatial extensions of a few thousand AU. In a turbulent environment these outflows are clearly asymmetric and less collimated than in the laminar case. Furthermore the outflows are ``better-shaped'' in the more magnetized cases, where the field is more ordered. The mean outflow velocity (mass-averaged) in the turbulent cases is about 0.5~km\ s$^{-1}$, while peak velocities are $\sim5$ km s$^{-1}$. In general, outflow velocities in the turbulent cases are about one order of magnitude lower than in the non-turbulent case.

Figure~\ref{img:Mout} shows the mass of the outflows for different magnetization parameters. Mass ejection rates are roughly constant $\sim10^{-5}M_{\odot}$\ yr$^{-1}$ for $\mu=2$ and 5 with turbulence, and about a few $10^{-5}$ to few $10^{-4}M_{\odot}$\ yr$^{-1}$ for $\mu=17$. The total mass in the outflows can be as large as a few $\times10^{-1}M_{\odot}$, indicating that a significant fraction of the mass of the first core is actually expelled. Indeed, as shown in Fig.~\ref{img:MoutvsMcore}, the ratio of the mass ejected to that accreted onto the first core (defined as the mass with a density higher than $10^{10}\,{\rm cm}^{-3}$) is between 0.3 to 0.7. This ratio tends to be larger with decreasing turbulent energy and magnetic field strength, but remains clearly of the same order. 

The fact that this ratio stays of the same order of magnitude is in good agreement with the work of \cite{Matzner00}, who showed that the efficiency of star formation ($\epsilon\sim M_{*}/M_{{\rm ej}}$, with $M_{*}$ the mass of the star and $M_{{\rm ej}}$ the mass in the outflows) is independent of the mass of the star. This has important consequence on the relation between the core mass function (CMF) and the initial mass function (IMF). If the star formation efficiency does not depend on the mass of the core, the CMF will naturally evolve into the IMF with a factor $\sim\epsilon$. This is therefore a plausible explanation of the evolution of the CMF into the IMF. We note however that we do not find the same numerical ratio \citep[between $\sim$1 and 3 instead of $\sim$1/3 and 1/2 in][]{Matzner00}. This is probably because we do not quantify exactly the same ratio, as we do not know the final mass of the star. We also emphasize that after the formation of the protostar, high velocity jets would be triggered, transporting more mass, and therefore decreasing the computed ratio.

\section{Discussions}
  \subsection{Turbulent magnetic diffusivity or misalignment?}

To quantify more precisely the influence of an initial misalignment angle, $\alpha$, between the rotation axis and the magnetic field on the transport angular momentum, and to disentangle its effects from those linked to turbulent diffusivity, we present here a series of simulations without turbulence, but with $\alpha\neq0$. In particular we take $\mu=5$, and choose $\alpha=57^{\circ}$ and $\alpha=80^{\circ}$, which correspond to the typical misalignment angles obtained in the simulations with turbulence. We stress however that for turbulent simulations it is difficult to clearly define an orientation angle (\emph{cf.} \ref{sec:orientation}). In addition, we remind that the results presented earlier for the cases without turbulence, denoted \eturb{} = 0, corresponded to an initially aligned core configuration (\emph{i.e.} $\alpha=0^{\circ}$ ).

\begin{figure*}
\includegraphics[width=1\textwidth]{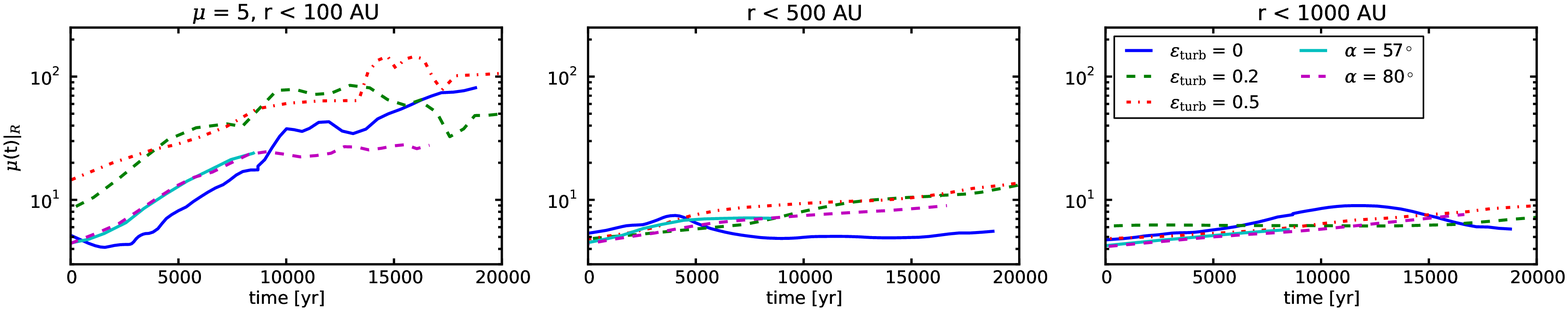}
\caption{Time evolution of the magnetization parameter within a radius $R=$ 100, 500 and 1000~AU around the central first core for an initial $\mu=5$. The cases shown include turbulence ($\eturb{}=0.2$ and 0.5), no turbulence and aligned ($\eturb{}=0$ and $\alpha=0^{\circ}$), and no turbulence but an initial misalignment.}  
\label{img:diffmucomp} 
\end{figure*}
 
Figure~\ref{img:diffmucomp} shows the evolution of the magnetization parameter $\mu(t)$ given by Eq.~\ref{eq:mu} within a radius $R=$ 50, 100, 500 and 1000~AU around the central first core. The magnetization parameter increases slightly more rapidly in the misaligned case than in the \eturb{} = 0, $\alpha=0^{\circ}$ case, but remains significantly lower than the turbulent cases. These results show that the diffusion of the magnetic field is not greatly affected by misalignment in the vicinity of the first core, this is expected since it is the turbulence that enhances the magnetic diffusivity.

\begin{figure}
\includegraphics[width=0.5\textwidth]{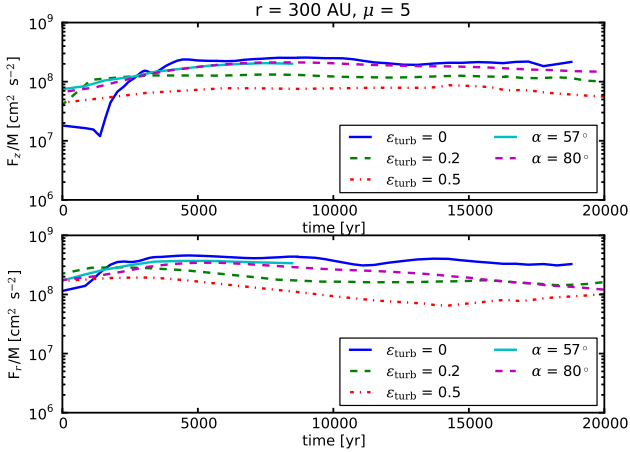}
\caption{Time evolution of the vertical and radial components of the flux of specific angular momentum transported by the magnetic field in logarithmic scale. The cases shown are for $\mu=5$, and include turbulence ($\eturb{}=0.2$ and 0.5), no turbulence and aligned ($\eturb{}=0$ and $\alpha=0^{\circ}$), and no turbulence but an initial misalignment ($\alpha=57^{\circ}$ and $\alpha=80^{\circ}$).}  
\label{img:brakcomp} 
\end{figure}
 
Figure~\ref{img:brakcomp} shows the evolution of $F_z$ and $F_r$ defined by Eq.~\ref{eq:magbrakcomp}. After a few thousand years these fluxes are lower for the misaligned cases than in the the \eturb{} = 0, $\alpha=0^{\circ}$ case, which indicates that the magnetic braking is less efficient with an initial misalignment, a result also shown in \cite{Joos12}. Compared to the laminar, aligned case, the magnetic braking components also grow more rapidly due to the growth of $B_r$ and $B_{\phi}$. It is also clear that $F_r$ and $F_z$ are larger for the cases with no-turbulence and an initial misalignment than with turbulence. This indicates that while the weaker magnetic braking observed in the turbulent cases is partly due to the initial misalignment induced by the turbulent velocity field, this alone is not enough to account for the whole effects of turbulence.

\begin{figure}
\includegraphics[width=0.5\textwidth]{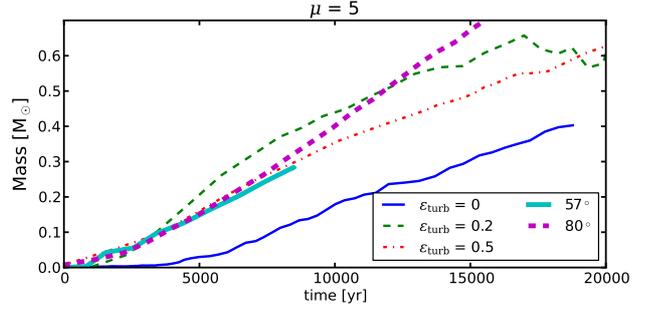}
\caption{Mass of the disk as a function of time. The cases shown are for $\mu=5$, and include turbulence ($\eturb{}=0.2$ and 0.5), no turbulence and aligned ($\eturb{}=0$ and $\alpha=0^{\circ}$), and no turbulence but an initial misalignment ($\alpha=57^{\circ}$ and $\alpha=80^{\circ}$) which are shown with thick lines.}
\label{img:diskcomp} 
\end{figure}
 
The effects of misalignment on the mass of the disks are shown in Figure \ref{img:diskcomp}. In first cores with an initial misalignment (and no turbulence), the mass of the disks are very comparable to those obtained with turbulence (\eturb{} = 0.2 and 0.5). In particular, the $\alpha=57^{\circ},\,\eturb{}=0$ case fits very well the simulation with \eturb{} = 0.5, with the disk mass growing to 0.3~$M_{\odot}$ about 10 000~yr after the formation of the first core. The $\alpha=80^{\circ},\,\eturb{}=0$ is also in good agreement with the simulation with \eturb{} = 0.2, the mass growing to 0.5~$M_{\odot}$ about 12 000~yr after the formation of the first core. Thus, the tilt of the rotation axis with respect to the magnetic field plays an important role on the formation of massive disks in the turbulent cores, although as we discussed earlier, turbulence induced magnetic diffusion is also playing a role as well. We suggest that a similar situation is also likely to occur in the simulations with turbulence carried out by \cite{Seifried12}. Although they reported strong misalignment's (between 60 and 180$^{\circ}$, see their Fig.~2, bottom panel), the formation of massive disks was attributed to the weak build-up of a toroidal magnetic field component. However, we find that independently of the magnetization and level of turbulence, the observed launching of outflows is a clear signature of the presence of a significant toroidal magnetic field component \citep[see also][]{Hennebelle08a,Ciardi10}, and as demonstrated here, we attribute massive disk formation to the combined effects of turbulent diffusion and misalignment. 

\subsection{The effects of different realisations of the turbulent velocity field}  \label{sec:seed}

\begin{figure}
  \subfigure[\label{img:am0s1}]{\includegraphics[width=.4\textwidth]{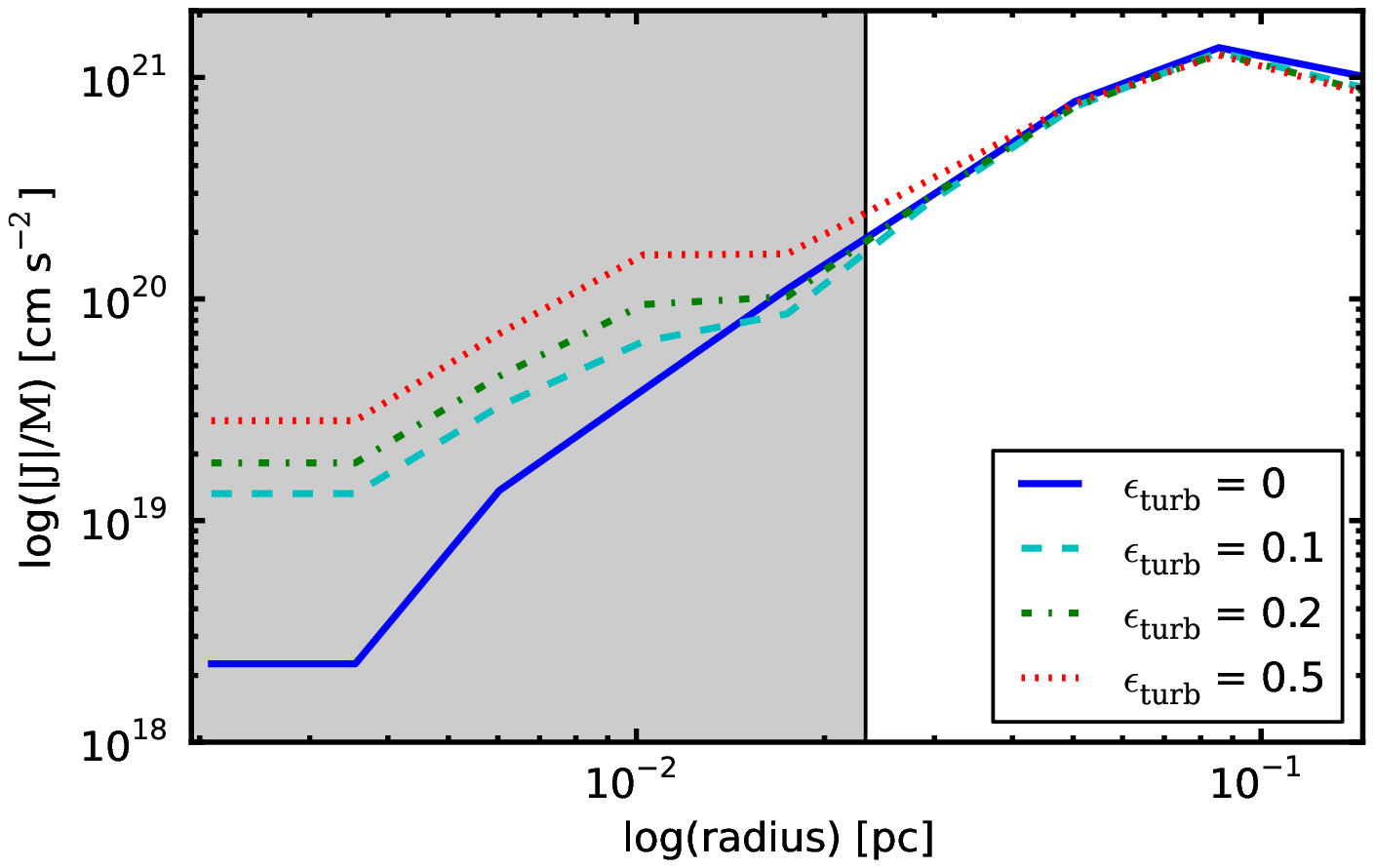}}
  \subfigure[\label{img:ang0s1}]{\includegraphics[width=.4\textwidth]{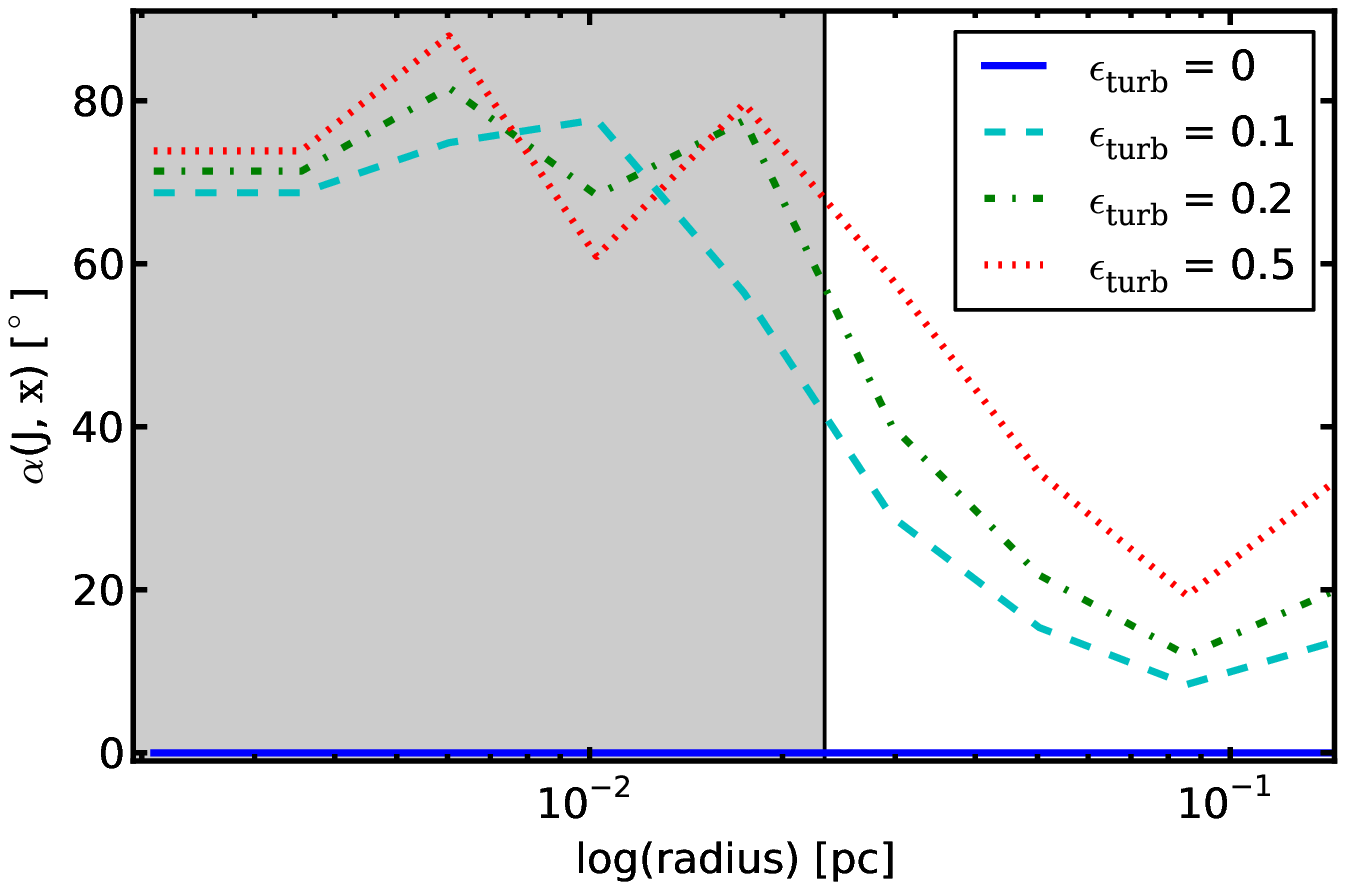}}
\caption{Specific angular momentum (upper panel) and initial angle between the rotation axis and the $x$-axis at $t=0$~yr, for the second realisation of the turbulent velocity field. The shaded area corresponds to the central region of the prestellar core of radius $r_0$.}  
\label{img:turbiniseed1} 
\end{figure} 

\begin{figure}
  \subfigure[\label{img:am0s2}]{\includegraphics[width=.4\textwidth]{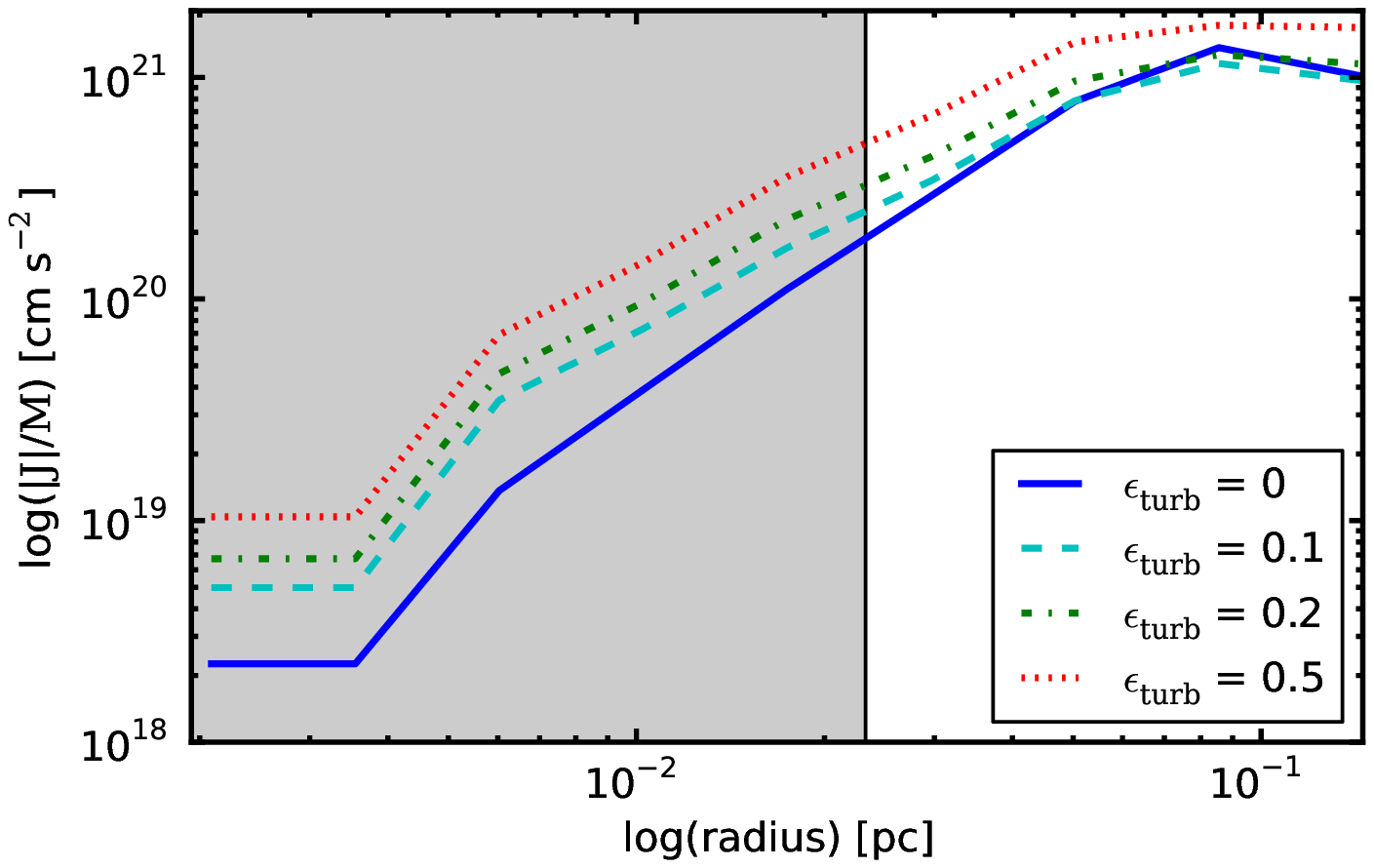}}
  \subfigure[\label{img:ang0s2}]{\includegraphics[width=.4\textwidth]{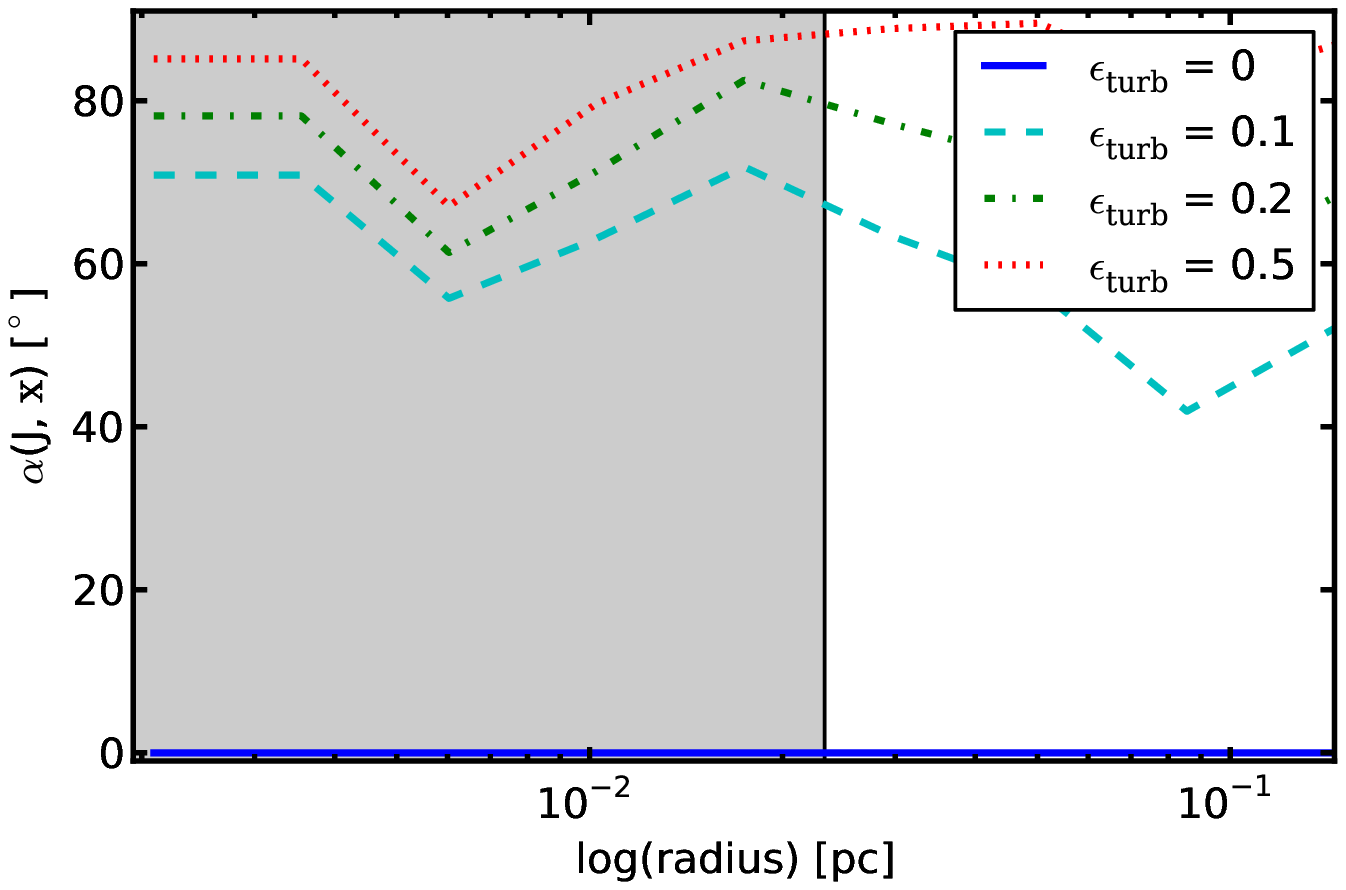}}
\caption{Specific angular momentum (upper panel) and initial angle between the rotation axis and the $x$-axis at $t=0$~yr, for the third realisation of the turbulent velocity field. The shaded area corresponds to the central region of the prestellar core of radius $r_0$.} 
\label{img:turbiniseed2} 
\end{figure}
 
We investigate here the influence of different realisations of the turbulent velocity field on the formation of disks and fragmentation. These realisations are denoted seed1 and seed2, whereas the first realisation is denoted seed0. We perform the simulations for magnetizations $\mu=5$ and 2, and for \eturb{} = 0.2 and 0.5. Since the turbulent velocity field has a non zero angular momentum, these different realisations of the velocity field have different angular momenta as shown in Fig.~\ref{img:turbiniseed1} and \ref{img:turbiniseed2}. We note in particular that for the second realisation of the turbulent velocity field (Fig.~\ref{img:turbiniseed1}), the angular momentum in the sphere of radius $r_0$ (corresponding to the radius of the central region of the prestellar core where the density is approximately constant), is the same, independently of the turbulent energy. Therefore, the angle between the rotation axis and the magnetic field is lower outside the prestellar core in this case. For the third realisation of the turbulent velocity field, the angular momentum and the initial angle are comparable to the ones obtained with the first realisation of the turbulent velocity field discussed previously.

The evolution of the magnetization for the three different realisations indicates that the diffusion of the magnetic field changes with the turbulent initial velocity field. For example, for $\mu=5$, it is initially more important with the first realisation; within 100~AU, the magnetization is already twice its initial value when the first core forms. For $\mu=2$, the diffusion also varies with the initial velocity field: for \eturb{} = 0.2 it is initially comparable, independently of the seed, and it is less efficient in removing magnetic flux from the first core regions for the first realisation with \eturb{} = 0.5. The angle between the rotation axis and the magnetic field also depends on the initial turbulent velocity field, and for the realisations explored it is $\alpha \gtrsim 20^{\circ}$\citep[which is the minimum angle required to form massive disks for $\mu = 5$,][]{Joos12}.

\begin{figure*}
  \subfigure[\label{img:diskseed5}]{\includegraphics[width=1.\textwidth]{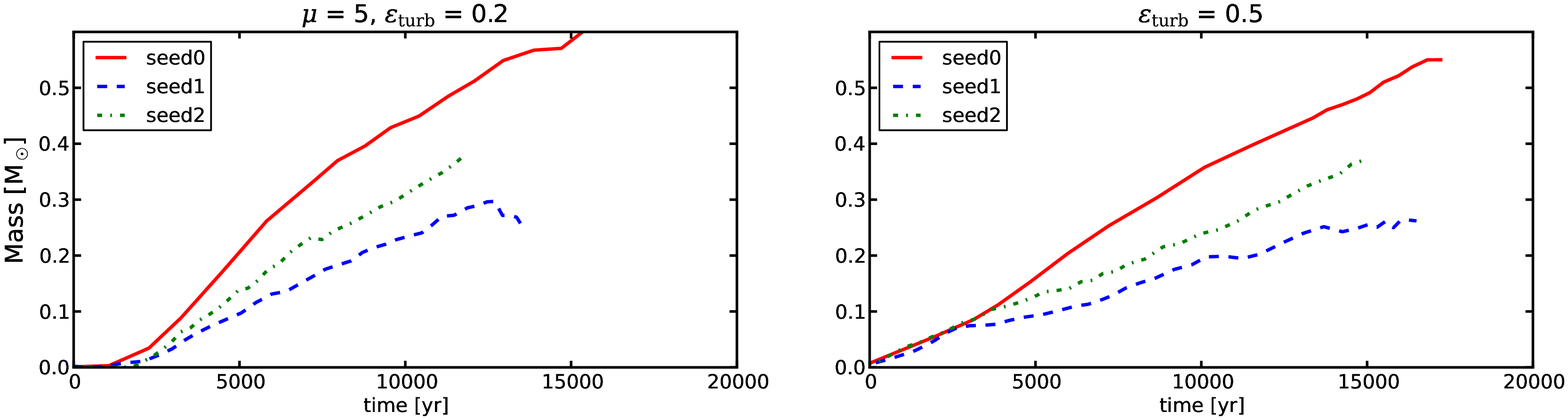}}
  \subfigure[\label{img:diskseed2}]{\includegraphics[width=1.\textwidth]{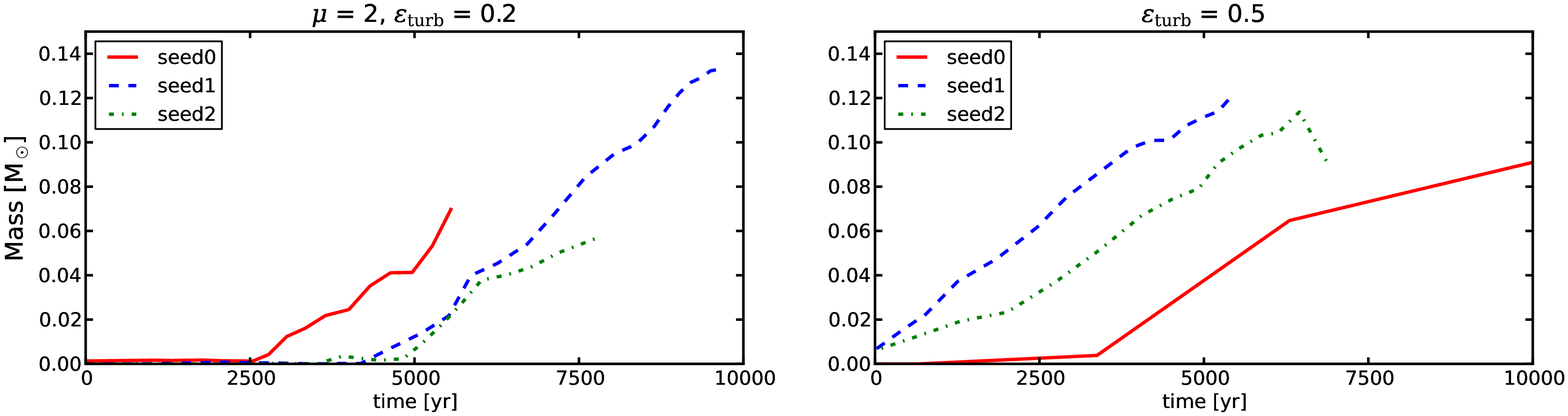}}
\caption{Mass of the disk as a function of time for $\mu=5$ and 2, for three different realisations, with \eturb{} = 0.2 and 0.5 (respectively left and right panel). The three realisations are denoted 'seed0' (solid lines), 'seed1' (dashed lines) and 'seed2' (dotted-dashed lines).}  
\label{img:diskseed} 
\end{figure*} 

The effects of different turbulent realisation on the mass of disks is presented in Figure~\ref{img:diskseed}. For $\mu=5$ (Fig.~\ref{img:diskseed5}), massive disks always form, but their mass can vary by a factor of the order of 2. At $t \sim$ 12 000~yr after the formation of the first core, this mass is between 0.3 and 0.6~$M_{\odot}$ with \eturb{} = 0.2, and between 0.2 and 0.5~$M_{\odot}$ with \eturb{} = 0.5. This is partly due to the inclination angle, and the turbulent magnetic diffusion, since the magnetic field is diffused more efficiently with the first realisation than with the two others. This is also coherent with the fact that the initial angular momentum was lower in the second realisation of the turbulent velocity field than in the other realisations. For $\mu=2$, $\eturb=0.5$, the trend is different than in the other cases: the disk is more massive with the second realisation and less massive with the first realisation of the turbulent velocity field. We believe the explanation is that the protostar forms in the central region of the cloud in the case seed1 were it can accrete more mass, whereas due to both the strong turbulence and magnetic support, it forms in outer regions of the cloud in the cases seed0 and seed2.

\section{Summary and conclusions}

We have carried out simulations of the collapse of turbulent prestellar dense cores of $5\, M_{\odot}$ with various magnetization parameters $\mu$, and presented a detailed analysis of the effects of turbulence in the star formation process. In particular, the influence of an effective turbulent magnetic diffusivity, and of the misalignment between the rotation axis and the magnetic field, on magnetic braking. In general, the turbulent velocity field imposed as initial conditions has important consequences on star formation, in particular regarding the formation of massive disks, their fragmentation and the launching of bipolar outflows. We also discussed the influence of the relative orientation of the rotation and field, in order to compare its effects to those of a turbulent velocity field, and, as a first step of a more complete statistical analysis, we performed simulations with two other realisations of the turbulent velocity field.

Our main conclusions are the following: 
\begin{itemize}
\item Magnetic braking is significantly reduced with turbulence for two reasons. First, turbulence efficiently diffuses the magnetic field out of the inner regions of the core. The turbulent magnetic diffusivity is responsible for a decrease of the magnetic field strength, and therefore of the magnetic braking within the core. Second, the turbulent velocity field contains a non-zero angular momentum which is responsible for a misalignment between the rotation axis and the magnetic field. As discussed here (and more thoroughly in \cite{Joos12}), this reduce significantly the magnetic braking. To quantify the effects of misalignment we compared the turbulent results with laminar cases having an initial misalignment. While this clearly plays an important role in the transport of angular momentum, the formation of disks and their fragmentation, it is not sufficient to account for all the effects of an initial turbulent velocity field. 
\item Owing to the reduced magnetic braking, massive disks can form more easily with turbulence than without. We study disk formation using the criteria defined in \cite{Joos12}, and we show that disks as massive as 1~$M_{\odot}$ can form after 25~000~yr with turbulence (in the $\mu=5$ case), whereas the mass of the disk formed in the case without turbulence this mass is about 0.5~$M_{\odot}$ at a comparable time. For $\mu=2$, the mass of the disks is of the order of 0.4~$M_{\odot}$, which is also twice the mass of the disks formed in the laminar case. Due to numerical convergence issue, these masses could however be overestimated, in particular in the more magnetized case (see the discussion in the appendix). As more massive disks are able to form, they will be more prone to fragmentation. Indeed, the more massive disk fragment into a few clumps, with masses growing rapidly from $\sim10^{-4}$ to $10^{-1}$~$M_{\odot}$ and then merge with the central object. We stress however that radiative feedback is not treated here and could modify the picture.
\item Slow outflows are launched in all simulations. In general turbulence and the tilt of the rotation axis tend to reduce the outflows strength compared to the laminar (aligned) case. In turbulent simulations the outflows in the higher magnetization cases ($\mu=2$) can be stronger than those ejected in the lower magnetization cases. In general, the mass of the outflows is a fraction of the mass of the protostellar core, and it has a weak dependence on the turbulent energy. 
\item A total of three different realisations of the initial turbulent velocity field were studied for the cases $\mu=5$ and $\mu=2$. The results are qualitatively similar: turbulence is responsible for an effective magnetic diffusion and misalignment, which reduce the magnetic braking and allow more massive disks to form. However, the results are quantitatively different, with the various initial turbulent velocity fields giving disks with masses within a factor$\sim$2 of each other. 
\item Regarding the magnetic diffusion and the magnetic braking, cores with subsonic turbulence are an intermediate case between the laminar and supersonic cases. However, the mass of the disks built in a subsonic environment are comparable to those formed in a supersonic environment. Our analysis is therefore robust regarding the turbulent environment of low-mass forming stars. 
\end{itemize}

{\small \emph{Acknowledgements.} We thank the anonymous referee for his/her thorough reading of the manuscript and his/her helpful comments and suggestions, which helped to significantly improve the quality of this article.}

\appendix
\section{Convergence} \label{sec:conv}

We run an additional set of simulations to investigate the convergence of our simulations. We change the Jeans refinement strategy to increase the spatial resolution. A cell was previously refined if its size exceeded one-tenth of a Jeans' length ($c_s (\pi/G \rho)^{1/2}$). We run simulations with 15 cells per Jeans' length (HR1) and another one with 20 cells per Jeans' length (HR2). We run this convergence run for $\mu = 2$, for \eturb{} = 0.2 and 0.5.

\subsection{Disk}

\begin{figure}
  \subfigure[\label{img:diskconv220}]{\includegraphics[width=.5\textwidth]{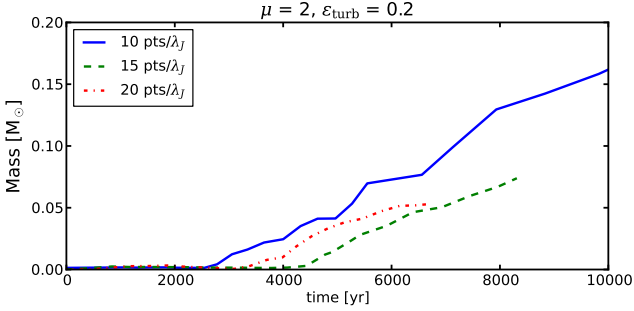}}
  \subfigure[\label{img:diskconv250}]{\includegraphics[width=.5\textwidth]{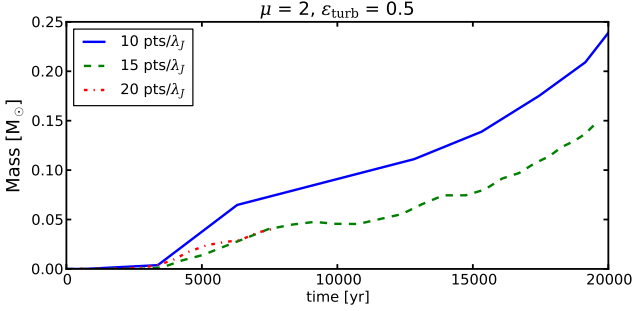}}
  \caption{Mass of the disk as a function of time for $\mu = 2$, for \eturb{} = 0.2 and 0.5 (respectively upper and lower panel), with 10, 15, and 20 points resolved per Jeans' length.}
  \label{img:diskconv}
\end{figure}

Figure~\ref{img:diskconv} shows the mass of the disk for $\mu = 2$, for \eturb{} = 0.2 and 0.5, with 10, 15, and 20 cells resolved per Jeans' length. It shows that convergence is not fully achieved in our simulations. Indeed, the disk is less massive by a factor up to $\sim$2 in the higher resolution runs.

\subsection{Outflows}

\begin{figure}
  \subfigure[\label{img:outconv220}]{\includegraphics[width=.5\textwidth]{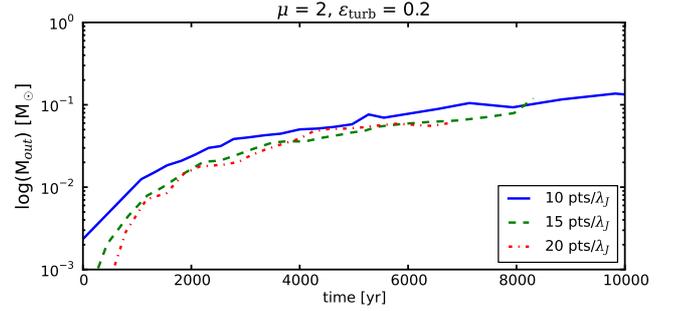}}
  \subfigure[\label{img:outconv250}]{\includegraphics[width=.5\textwidth]{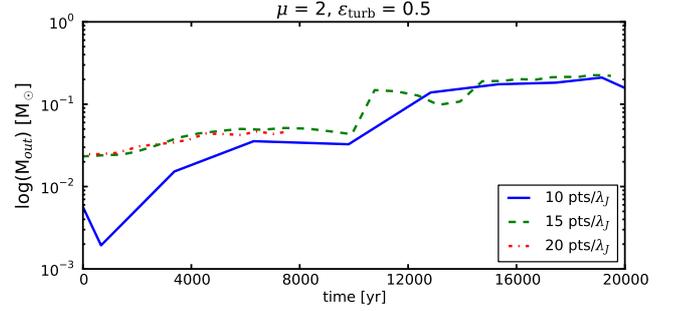}}
  \caption{Total mass in the outflows as a function of time for $\mu = 2$, for \eturb{} = 0.2 and 0.5, with 10, 15, and 20 points resolved per Jeans' length.}
  \label{img:outconv}
\end{figure}

Figure~\ref{img:outconv} shows the mass transported in the outflows for $\mu = 2$, for \eturb{} = 0.2 and 0.5, with 10, 15, and 20 cells resolved per Jeans' length. In the low resolution simulation, the outflows transport roughly twice the mass transported in the high resolutions simulations.
\newline

Even if our conclusions are qualitatively unchanged, these convergence runs show that our simulations have not converged yet, and that we overestimate the mass in the disk or in the outflows by a factor two for the $\mu = 2$ simulations. It is partly due to the fact that the disks are small on these simulations; it should therefore affect less strongly the simulations with a lower magnetization.

\bibliographystyle{aa}
\nocite{*}
\bibliography{bib_turb}

\begin{thebibliography}{55}
\expandafter\ifx\csname natexlab\endcsname\relax\def\natexlab#1{#1}\fi

\bibitem[{{Andr\'e} {et~al.}(2000){Andr\'e}, {Ward-Thompson}, \&
  {Barsony}}]{Andre00}
{Andr\'e}, P., {Ward-Thompson}, D., \& {Barsony}, M. 2000, Protostars and
  Planets IV, 59

\bibitem[{{Bacciotti} {et~al.}(2002){Bacciotti}, {Ray}, {Mundt},
  {Eisl{\"o}ffel}, \& {Solf}}]{Bacciotti02}
{Bacciotti}, F., {Ray}, T.~P., {Mundt}, R., {Eisl{\"o}ffel}, J., \& {Solf}, J.
  2002, ApJ, 576, 222

\bibitem[{{Banerjee} {et~al.}(2006){Banerjee}, {Pudritz}, \&
  {Anderson}}]{Banerjee06}
{Banerjee}, R., {Pudritz}, R.~E., \& {Anderson}, D.~W. 2006, MNRAS, 373, 1091

\bibitem[{{Bate} {et~al.}(2002){Bate}, {Bonnell}, \& {Bromm}}]{Bate02}
{Bate}, M.~R., {Bonnell}, I.~A., \& {Bromm}, V. 2002, MNRAS, 336, 705

\bibitem[{{Bate} {et~al.}(1995){Bate}, {Bonnell}, \& {Price}}]{Bate95}
{Bate}, M.~R., {Bonnell}, I.~A., \& {Price}, N.~M. 1995, MNRAS, 277, 362

\bibitem[{{Belloche} {et~al.}(2002){Belloche}, {Andr{\'e}}, {Despois}, \&
  {Blinder}}]{Belloche02}
{Belloche}, A., {Andr{\'e}}, P., {Despois}, D., \& {Blinder}, S. 2002, A\&A,
  393, 927

\bibitem[{{Blandford} \& {Payne}(1982)}]{Blandford82}
{Blandford}, R.~D. \& {Payne}, D.~G. 1982, MNRAS, 199, 883

\bibitem[{{Bonnell} {et~al.}(1991){Bonnell}, {Martel}, {Bastien}, {Arcoragi},
  \& {Benz}}]{Bonnell91}
{Bonnell}, I., {Martel}, H., {Bastien}, P., {Arcoragi}, J.-P., \& {Benz}, W.
  1991, ApJ, 377, 553

\bibitem[{{Boss} \& {Bodenheimer}(1979)}]{Boss79}
{Boss}, A.~P. \& {Bodenheimer}, P. 1979, ApJ, 234, 289

\bibitem[{{Ciardi} \& {Hennebelle}(2010)}]{Ciardi10}
{Ciardi}, A. \& {Hennebelle}, P. 2010, MNRAS, 409, L39

\bibitem[{{Commer{\c c}on} {et~al.}(2010){Commer{\c c}on}, {Hennebelle},
  {Audit}, {Chabrier}, \& {Teyssier}}]{Commercon10}
{Commer{\c c}on}, B., {Hennebelle}, P., {Audit}, E., {Chabrier}, G., \&
  {Teyssier}, R. 2010, A\&A, 510, L3

\bibitem[{{Commer{\c c}on} {et~al.}(2011{\natexlab{a}}){Commer{\c c}on},
  {Hennebelle}, \& {Henning}}]{Commercon11a}
{Commer{\c c}on}, B., {Hennebelle}, P., \& {Henning}, T. 2011{\natexlab{a}},
  ApJL, 742, L9

\bibitem[{{Commer{\c c}on} {et~al.}(2011{\natexlab{b}}){Commer{\c c}on},
  {Teyssier}, {Audit}, {Hennebelle}, \& {Chabrier}}]{Commercon11b}
{Commer{\c c}on}, B., {Teyssier}, R., {Audit}, E., {Hennebelle}, P., \&
  {Chabrier}, G. 2011{\natexlab{b}}, A\&A, 529, A35

\bibitem[{{Crutcher}(1999)}]{Crutcher99}
{Crutcher}, R.~M. 1999, ApJ, 520, 706

\bibitem[{{Dapp} \& {Basu}(2010)}]{Dapp10}
{Dapp}, W. \& {Basu}, S. 2010, A\&A, 521, 56

\bibitem[{{Duquennoy} \& {Mayor}(1991)}]{Duquennoy91}
{Duquennoy}, A. \& {Mayor}, M. 1991, A\&A, 248, 485

\bibitem[{{Falgarone} {et~al.}(2008){Falgarone}, {Troland}, {Crutcher}, \&
  {Paubert}}]{Falgarone08}
{Falgarone}, E., {Troland}, T.~H., {Crutcher}, R.~M., \& {Paubert}, G. 2008,
  A\&A, 487, 247

\bibitem[{{Ferreira}(1997)}]{Ferreira97}
{Ferreira}, J. 1997, A\&A, 319, 340

\bibitem[{{Fromang} {et~al.}(2006){Fromang}, {Hennebelle}, \&
  {Teyssier}}]{Fromang06a}
{Fromang}, S., {Hennebelle}, P., \& {Teyssier}, R. 2006, A\&A, 457, 371

\bibitem[{{Fromang} \& {Papaloizou}(2006)}]{Fromang06}
{Fromang}, S. \& {Papaloizou}, J. 2006, A\&A, 452, 751

\bibitem[{{Goodwin} {et~al.}(2004{\natexlab{a}}){Goodwin}, {Whitworth}, \&
  {Ward-Thompson}}]{Goodwin04a}
{Goodwin}, S.~P., {Whitworth}, A.~P., \& {Ward-Thompson}, D.
  2004{\natexlab{a}}, A\&A, 414, 633

\bibitem[{{Goodwin} {et~al.}(2004{\natexlab{b}}){Goodwin}, {Whitworth}, \&
  {Ward-Thompson}}]{Goodwin04b}
{Goodwin}, S.~P., {Whitworth}, A.~P., \& {Ward-Thompson}, D.
  2004{\natexlab{b}}, A\&A, 423, 169

\bibitem[{{Hennebelle} \& {Ciardi}(2009)}]{Hennebelle09}
{Hennebelle}, P. \& {Ciardi}, A. 2009, A\&A, 506, L29

\bibitem[{{Hennebelle} {et~al.}(2011){Hennebelle}, {Commer{\c c}on}, {Joos},
  {Klessen}, {Krumholz}, {Tan}, \& {Teyssier}}]{Hennebelle11}
{Hennebelle}, P., {Commer{\c c}on}, B., {Joos}, M., {et~al.} 2011, A\&A, 528,
  A72+

\bibitem[{{Hennebelle} \& {Fromang}(2008)}]{Hennebelle08a}
{Hennebelle}, P. \& {Fromang}, S. 2008, A\&A, 477, 9

\bibitem[{{Hennebelle} \& {Teyssier}(2008)}]{Hennebelle08b}
{Hennebelle}, P. \& {Teyssier}, R. 2008, A\&A, 477, 25

\bibitem[{{Joos} {et~al.}(2012){Joos}, {Hennebelle}, \& {Ciardi}}]{Joos12}
{Joos}, M., {Hennebelle}, P., \& {Ciardi}, A. 2012, A\&A, (in press)

\bibitem[{{Krasnopolsky} {et~al.}(2010){Krasnopolsky}, {Li}, \&
  {Shang}}]{Krasnopolsky10}
{Krasnopolsky}, R., {Li}, Z.-Y., \& {Shang}, H. 2010, ApJ, 716, 1541

\bibitem[{{Krasnopolsky} {et~al.}(2012){Krasnopolsky}, {Li}, {Shang}, \&
  {Zhao}}]{Krasnopolsky12}
{Krasnopolsky}, R., {Li}, Z.-Y., {Shang}, H., \& {Zhao}, B. 2012, ApJ, 757, 77

\bibitem[{{Larson}(1981)}]{Larson81}
{Larson}, R.~B. 1981, MNRAS, 194, 809

\bibitem[{{Larson}(2003)}]{Larson03}
{Larson}, R.~B. 2003, Reports on Progress in Physics, 66, 1651

\bibitem[{{Levrier}(2004)}]{Levrier04}
{Levrier}, F. 2004, PhD thesis, Universit\'e Paris VII

\bibitem[{{Li} {et~al.}(2011){Li}, {Krasnopolsky}, \& {Shang}}]{Li11}
{Li}, Z.-Y., {Krasnopolsky}, R., \& {Shang}, H. 2011, ApJ, 738, 180

\bibitem[{{Lissauer}(1993)}]{Lissauer93}
{Lissauer}, J.~J. 1993, ARAA, 31, 129

\bibitem[{{Machida} {et~al.}(2008){Machida}, {Inutsuka}, \&
  {Matsumoto}}]{Machida08}
{Machida}, M.~N., {Inutsuka}, S.-i., \& {Matsumoto}, T. 2008, ApJ, 676, 1088

\bibitem[{{Machida} {et~al.}(2005){Machida}, {Matsumoto}, {Hanawa}, \&
  {Tomisaka}}]{Machida05}
{Machida}, M.~N., {Matsumoto}, T., {Hanawa}, T., \& {Tomisaka}, K. 2005, MNRAS,
  362, 382

\bibitem[{{Matsumoto} \& {Hanawa}(2011)}]{Matsumoto11}
{Matsumoto}, T. \& {Hanawa}, T. 2011, ApJ, 728, 47

\bibitem[{{Matzner} \& {McKee}(2000)}]{Matzner00}
{Matzner}, C.~D. \& {McKee}, C.~F. 2000, ApJ, 545, 364

\bibitem[{{Maury} {et~al.}(2010){Maury}, {Andr{\'e}}, {Hennebelle}, {Motte},
  {Stamatellos}, {Bate}, {Belloche}, {Duch{\^e}ne}, \& {Whitworth}}]{Maury10}
{Maury}, A.~J., {Andr{\'e}}, P., {Hennebelle}, P., {et~al.} 2010, A\&A, 512,
  A40+

\bibitem[{{Mellon} \& {Li}(2008)}]{Mellon08}
{Mellon}, R.~R. \& {Li}, Z. 2008, ApJ, 681, 1356

\bibitem[{{Mellon} \& {Li}(2009)}]{Mellon09}
{Mellon}, R.~R. \& {Li}, Z. 2009, ApJ, 698, 922

\bibitem[{{Miyoshi} \& {Kusano}(2005)}]{Miyoshi05}
{Miyoshi}, T. \& {Kusano}, K. 2005, JCoPh, 208, 315

\bibitem[{{Offner} {et~al.}(2009){Offner}, {Klein}, {McKee}, \&
  {Krumholz}}]{Offner09}
{Offner}, S.~S.~R., {Klein}, R.~I., {McKee}, C.~F., \& {Krumholz}, M.~R. 2009,
  ApJ, 703, 131

\bibitem[{{Offner} {et~al.}(2010){Offner}, {Kratter}, {Matzner}, {Krumholz}, \&
  {Klein}}]{Offner10}
{Offner}, S.~S.~R., {Kratter}, K.~M., {Matzner}, C.~D., {Krumholz}, M.~R., \&
  {Klein}, R.~I. 2010, ApJ, 725, 1485

\bibitem[{{Pelletier} \& {Pudritz}(1992)}]{Pelletier92}
{Pelletier}, G. \& {Pudritz}, R.~E. 1992, ApJ, 394, 117

\bibitem[{{Price} \& {Bate}(2007)}]{Price07}
{Price}, D.~J. \& {Bate}, M.~R. 2007, Ap\&SS, 311, 75

\bibitem[{{Pudritz} \& {Norman}(1983)}]{Pudritz83}
{Pudritz}, R.~E. \& {Norman}, C.~A. 1983, ApJ, 274, 677

\bibitem[{{Santos-Lima} {et~al.}(2012){Santos-Lima}, {de Gouveia Dal Pino}, \&
  {Lazarian}}]{Santos-Lima12}
{Santos-Lima}, R., {de Gouveia Dal Pino}, E.~M., \& {Lazarian}, A. 2012, ApJ,
  747, 21

\bibitem[{{Seifried} {et~al.}(2012){Seifried}, {Banerjee}, {Pudritz}, \&
  {Klessen}}]{Seifried12}
{Seifried}, D., {Banerjee}, R., {Pudritz}, R.~E., \& {Klessen}, R.~S. 2012,
  MNRAS, L442

\bibitem[{{Stamatellos} {et~al.}(2007){Stamatellos}, {Hubber}, \&
  {Whitworth}}]{Stamatellos07}
{Stamatellos}, D., {Hubber}, D.~A., \& {Whitworth}, A.~P. 2007, MNRAS, 382, L30

\bibitem[{{Teyssier}(2002)}]{Teyssier02}
{Teyssier}, R. 2002, A\&A, 385, 337

\bibitem[{{Tomida} {et~al.}(2010){Tomida}, {Machida}, {Saigo}, {Tomisaka}, \&
  {Matsumoto}}]{Tomida10}
{Tomida}, K., {Machida}, M.~N., {Saigo}, K., {Tomisaka}, K., \& {Matsumoto}, T.
  2010, ApJL, 725, L239

\bibitem[{{Uchida} \& {Shibata}(1985)}]{Uchida85}
{Uchida}, Y. \& {Shibata}, K. 1985, PASJ, 37, 515

\bibitem[{{Ward-Thompson} {et~al.}(2007){Ward-Thompson}, {Andr{\'e}},
  {Crutcher}, {Johnstone}, {Onishi}, \& {Wilson}}]{Ward-Thompson07}
{Ward-Thompson}, D., {Andr{\'e}}, P., {Crutcher}, R., {et~al.} 2007, Protostars
  and Planets V, 33

\bibitem[{{Weiss}(1966)}]{Weiss66}
{Weiss}, N.~O. 1966, Royal Society of London Proceedings Series A, 293, 310

\end{thebibliography}

\end{document}